\documentclass[11pt,draftcls,onecolumn]{IEEEtran}
\usepackage{color}
\usepackage{mdwlist}
\usepackage{subfigure}
\usepackage{setspace}
\usepackage{comment}
\usepackage{booktabs}
\usepackage{graphicx}
\usepackage{amsthm,amsfonts,amsmath,amssymb}
\usepackage[numbers]{natbib}
\usepackage[colorlinks=true,linkcolor=blue,citecolor=magenta]{hyperref}
\usepackage{multirow}
\usepackage{tabulary}
\usepackage{algorithm}
\usepackage{algorithmic}
\usepackage{enumerate}

\newcommand{\algrule}[1][1pt]{\par\vskip.5\baselineskip\hrule height #1\par\vskip.5\baselineskip}
\newcommand{\mbf}{\boldsymbol}

\newcommand{\nn}{\nonumber}
\newcommand{\eop}{\hfill $\blacksquare$}

\newcommand{\pr}{Pr}
\newcommand{\eye}[2]{{I}_{ {#1}\times{#2} } }
\newcommand{\zvector}{\vec{0}}
\newcommand{\complex}{\mathbb{C}}
\newcommand{\real}{\mathbb{R}}

\newcommand{\zeronorm}[1]{||{#1}||_0}

\newcommand{\expectation}{\mathbb{E}}

\newcommand{\constants}[1]{c_{#1}}

\newcommand{\threshold}{\gamma}
\newcommand{\range}{c_1}

\newcommand{\normal}{{\cal N}}
\newcommand{\cnormal}{{\cal CN}}
\newcommand{\noise}{\vec{z}}

\newcommand{\nbinobsv}[2]{\vec{w}_{b,#1,#2}}
\newcommand{\nbinobsvb}{\vec{w}_{b}}
\newcommand{\nbinobsc}[2]{w_{b,#1,#2}}

\newcommand{\levels}{{M}}
\newcommand{\snr}{\rho}
\newcommand{\signal}{\vec{x}}
\newcommand{\signalc}[1]{x[{#1}]}
\newcommand{\transform}{\vec{X}}
\newcommand{\transformc}[1]{X[{#1}]}

\newcommand{\mdft}{$2D$-DFT}

\newcommand{\obs}{\vec{y}}

\newcommand{\binobsv}[2]{\vec{y}_{b,#1,#2}}
\newcommand{\binobsvb}{\vec{y}_{b}}

\newcommand{\mmatrix}{\mbf{A}}
\newcommand{\svector}[1]{\vec{a}({#1})}
\newcommand{\coh}{\mu}
\newcommand{\maxcoh}{\coh_{\max}}
\newcommand{\ripc}{\gamma}

\newcommand{\stages}{d}

\newcommand{\factor}[1]{f_{{#1}}}
\newcommand{\nbins}{n_b}
\newcommand{\delays}{D}
\newcommand{\clusters}{C}
\newcommand{\samplesc}{N}
\newcommand{\constellation}{{\mathcal A}}
\newcommand{\magset}{{\mathcal M}}
\newcommand{\phaseset}{{\Theta}}

\newcommand{\eventsb}{E}
\newcommand{\eventf}{E_f}
\newcommand{\eventb}{E_b}
\newcommand{\eventz}{E_z}
\newcommand{\events}{E_s}
\newcommand{\eventsz}{E_{sz}}
\newcommand{\eventss}{E_{ss}}
\newcommand{\eventsm}{E_{sm}}
\newcommand{\eventm}{E_m}

\newcommand{\length}{n}
\newcommand{\sparsity}{k}
\newcommand{\sindex}{\delta}
\newcommand{\samples}{m}

\newcommand{\iterations}{\ell}
\newcommand{\binexp}{\eta} % |bin| = \eta k

\newcommand{\primef}[1]{{\cal P}_{#1}}

\newtheorem{theorem}{Theorem}[section]

\newtheorem{remark}[theorem]{Remark}%[section]
\newtheorem{prop}[theorem]{Proposition}
\newtheorem{definition}[theorem]{Definition}

\newtheorem{lemma}[theorem]{Lemma}
\newtheorem{thm}[theorem]{Theorem}

\begin{document}
\title{A robust sub-linear time R-FFAST algorithm for computing a sparse DFT}
\author{
\authorblockN{Sameer~Pawar~and~Kannan~Ramchandran\\}
\authorblockA{Dept. of Electrical Engineering and Computer Sciences \\
University of California, Berkeley \\
\{spawar, kannanr\}@eecs.berkeley.edu}
}

\maketitle
%*********************************************************************************%
% Abstract
%*********************************************************************************%
\begin{abstract}
The Fast Fourier Transform (FFT) is the most efficiently known way to compute the Discrete Fourier Transform (DFT) of an arbitrary $\length$-length signal, and has a computational complexity of $O(\length \log\length)$. If the DFT $\transform$ of the signal $\signal$ has only $k$ non-zero coefficients (where $\sparsity < \length$), can we do better? In \cite{pawar2013computing}, we addressed this question and presented a novel FFAST (Fast Fourier Aliasing-based Sparse Transform) algorithm that cleverly induces sparse graph alias codes in the DFT domain, via a Chinese-Remainder-Theorem (CRT)-guided sub-sampling operation of the time-domain samples. The resulting sparse graph alias codes are then exploited to devise a fast and iterative onion-peeling style decoder that computes an $\length$ length DFT of a signal using only $O(\sparsity)$ time-domain samples and $O(\sparsity\log\sparsity)$ computations. The FFAST algorithm is applicable whenever $\sparsity$ is sub-linear in $\length$ (i.e. $\sparsity = o(\length)$), but is obviously most attractive when $\sparsity$ is much smaller than $\length$.

In this paper, we adapt the FFAST framework of \cite{pawar2013computing} to the case where the time-domain samples are corrupted by a white Gaussian noise. In particular, we show that the extended noise robust algorithm R-FFAST computes an $\length$-length $\sparsity$-sparse DFT $\transform$ using $O(\sparsity\log^{3}\length)$\footnote{In this paper, we assume a {\em fixed} sampling structure or measurement system, that is motivated by practical constraints, particularly for hardware implementation.  If it is feasible to have a randomized measurement system, which permits the flexibility of choosing a different sampling structure for every input signal, we also have a variant of the R-FFAST algorithm presented in \cite{li2014sub} which can robustly compute an $\length$-length $\sparsity$-sparse DFT using $O(\sparsity\log^{4/3}\length)$ samples and $O(\sparsity\log^{7/3}\length)$ computations.} noise-corrupted time-domain samples, in $O(\sparsity\log^{4}\length)$ computations, i.e., {\em sub-linear time complexity}. While our theoretical results are for signals with a uniformly random support of the non-zero DFT coefficients and additive white Gaussian noise, we provide simulation results which demonstrates that the R-FFAST algorithm performs well even for signals like MR images, that have an approximately sparse Fourier spectrum with a non-uniform support for the dominant DFT coefficients.\end{abstract}
%*********************************************************************************%
% Introduction
%*********************************************************************************%
\setcounter{subsubsection}{0}
\section{Introduction}\label{sec:intro}

The Fast Fourier Transform is the fastest known way to compute the DFT of an arbitrary $\length$-length signal, and has a computational complexity of $O(\length \log\length)$. Many applications of interest involve signals that have a sparse Fourier spectrum, e.g. signals relating to audio, image, and video data, biomedical signals, etc. In \cite{pawar2013computing}, we have proposed a novel FFAST (Fast Fourier Aliasing-based Sparse Transform) framework, that under idealized assumptions of no measurement noise and an exactly $\sparsity$-sparse signal spectrum (i.e. there are precisely $\sparsity$ non-zero DFT coefficients), computes an $\length$-length DFT $\transform$ using only $O(\sparsity)$ time-domain samples in $O(\sparsity\log\sparsity)$ arithmetic computations. For large signals, i.e., when $\length$ is of order of millions or tens of millions, as is becoming more relevant in the Big-data age, the gains over  conventional FFT algorithms can be significant. The idealized assumptions in \cite{pawar2013computing} were primarily used to highlight the conceptual framework underlying the FFAST architecture. 

\begin{figure}[t]
\centering     
\subfigure[Log intensity plot of the \mdft \ of the original `Brain' image.]{\label{fig:dftfull}\includegraphics[width=.32\linewidth]{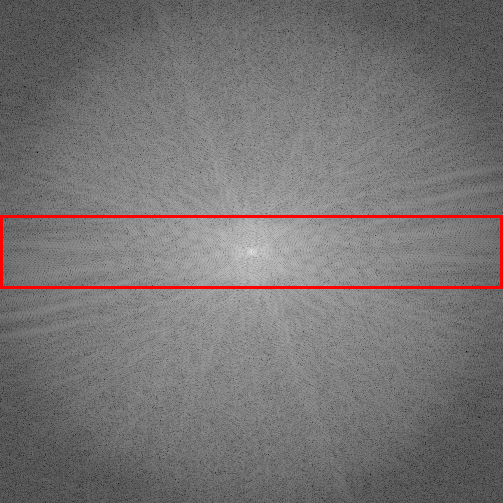}}
\subfigure[Original `Brain' image in spatial domain.]{\label{fig:spatialfull}\includegraphics[width=.32\linewidth]{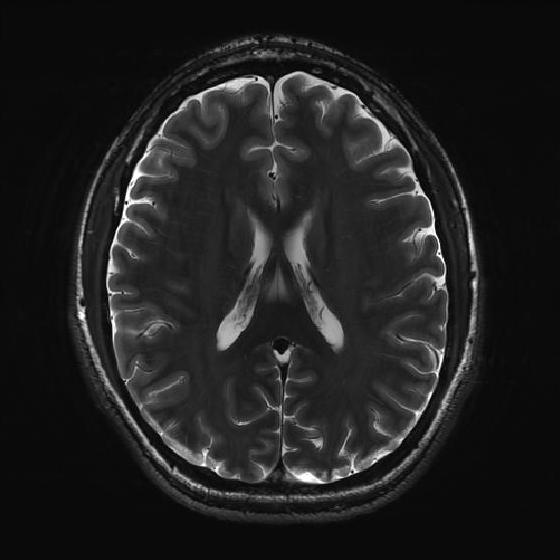}}
\subfigure[Reconstructed `Brain' image using the R-FFAST.]{\label{fig:rfull}\includegraphics[width=.32\linewidth]{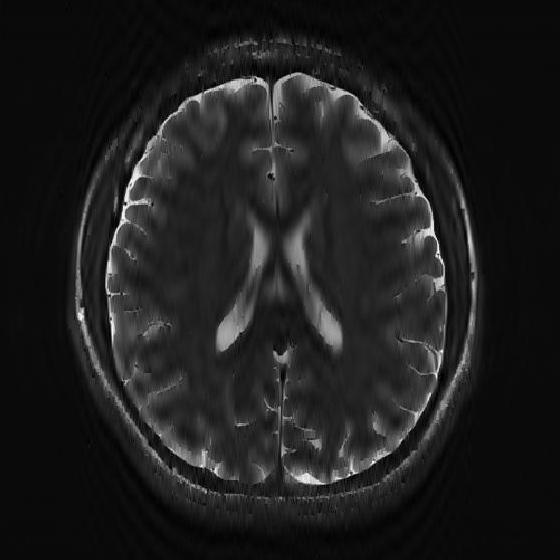}}
\caption{Application of the $2D$ R-FFAST algorithm to reconstruct the `Brain' image acquired on an MR scanner with dimension $504 \times 504$. The $2D$ R-FFAST algorithm reconstructs the `Brain' image, as shown in Fig.~\ref{fig:rfull}, using overall $60.18\%$ of the Fourier samples of Fig.~\ref{fig:dftfull}.}\vspace*{-0.2in}\label{fig:MRI}
\end{figure}

In this paper, we adapt the noiseless FFAST framework of \cite{pawar2013computing} to a noise robust R-FFAST algorithm. The robustness to measurement noise is achieved by judiciously modifying the FFAST framework, as explained in details in Section~\ref{sec:delayCS}. Specifically, the FFAST front-end in \cite{pawar2013computing} has multiple stages of sub-sampling operations, where each stage further has $2$ sub-sampling paths called the delay-chains. Both the delay-chains in each stage of the FFAST framework have an identical sampling period but a circularly time-shifted input signal, i.e., {\em consecutive shifts}. In contrast, in R-FFAST we use {\em a large number of random shifts}: more precisely, an asymptotically large (i.e. $O(\log^3 \length)$ number of delay chains per stage, where the input to each delay-chain is shifted by a random amount prior to being subsampled (see Section~\ref{sec:FFASTArch}). A random choice of the circular shifts endows the effective measurement matrix with good {\em mutual incoherence} and {\em Restricted Isometry properties (RIP)} \cite{candes2005decoding}, thus enabling stable recovery.

As a motivating example, consider an application of a $2D$ version of the R-FFAST algorithm to acquire the Magnetic Resonance Image (MRI) of the `Brain' as shown in Fig.~\ref{fig:MRI}. In MRI, recall that the samples are acquired in the Fourier domain, and the challenge is to speed up the acquisition time by minimizing the number of Fourier samples needed to reconstruct the desired spatial domain image. The R-FFAST algorithm reconstructs the `Brain' image acquired on an MR scanner from $60.18\%$ of the Fourier samples as shown in Fig.~\ref{fig:rfull}. In Section~\ref{sec:sim_brain}, we elaborate on the details of this experiment. These results are not meant to compete with the state-of-the-art techniques in the MRI literature but rather to demonstrate the feasibility of our proposed approach as a promising direction for future research.  More significantly, this experiment demonstrates that while the theory proposed in this paper applies to signals with a uniformly-random support  model for the dominant sparse DFT coefficients,  in practice, our algorithm works even for the non-uniform (or clustered) support setting as is typical of MR images. However, the focus of this paper is to design a robust sub-linear time algorithm to compute $\sparsity$-sparse $1D$ DFT of $\length$-length signals from its noise corrupted time-domain samples.

In many practical settings, particularly hardware-based applications, it is desirable or even required to have a {\em fixed} sampling structure (or measurement system) that has to be used for {\em all} input signals to be processed.  It is this setting that we focus on in this work. Under this setting, we show that the proposed R-FFAST algorithm computes an $\length$-length $\sparsity$-sparse DFT $\transform$ using $O(\sparsity\log^3\length)$ noise-corrupted time-domain samples, in $O(\sparsity\log^4\length)$ complex operations, i.e., {\em sub-linear sample and time complexity}. A variant of the R-FFAST algorithm presented in \cite{li2014sub} is useful for applications that have the flexibility of choosing a {\em different} measurement matrix for every input signal (i.e. permit a {\em randomized} measurement system).  Under this setting, which we do not elaborate on in this work, our algorithm can compute an $\length$-length $\sparsity$-sparse DFT using $O(\sparsity\log^{4/3}\length)$ samples and $O(\sparsity\log^{7/3}\length)$ computations. We emphasize the following caveats for the results in this paper. First, we assume that the non-zero DFT coefficients of the signal $\signal$ have uniformly random support and take values from a finite constellation (see Section~\ref{sec:model} for more details). Secondly, our results are probabilistic and are applicable for asymptotic values of $\sparsity, \length$, where $\sparsity$ is sub-linear in $\length$. Lastly, we assume an i.i.d Gaussian noise model for observation noise.

The rest of the paper is organized as follows: In Section~\ref{sec:model}, we provide the problem formulation and the signal model. Section~\ref{sec:main} provides the main result of this paper. An overview of the related literature is provided in Section~\ref{sec:related}. In Section~\ref{sec:ExFFAST}, we use a simple example to explain the key ingredients of the R-FFAST architecture and the algorithm. Section~\ref{sec:FreqEst}, provides a brief review of some Frequency-estimation results from signal processing literature, that are key to our proposed recovery algorithm. In Section~\ref{sec:FFASTArch} and Section~\ref{sec:backEnd}, we provide a generic description of both the R-FFAST architecture as well as the recovery algorithm. Section~\ref{sec:simulations} provides experimental results that validate the performance of the R-FFAST algorithm.

%*********************************************************************************%
% Problem Formulation
%*********************************************************************************%
\setcounter{subsubsection}{0}
\section{Signal model and Problem formulation}\label{sec:model}
Consider an $\length$-length discrete-time signal $\signal$ that is a sum of $\sparsity << \length$ complex exponentials, i.e., its $\length$-length discrete Fourier transform has $\sparsity$ non-zero coefficients:
\begin{equation}
\signalc{p} = \sum_{q = 0}^{\sparsity-1} \transformc{\ell_q} e^{2\pi \imath \ell_q p/\length}, \ \ \ \ p = 0,1,\hdots,\length-1,
\end{equation}
where the discrete frequencies $\ell_q \in \{0,1,\hdots,\length-1 \}$ and the amplitudes $\transformc{\ell_q}\in {\complex}$. We consider the problem of computing the $\sparsity$-sparse $\length$-length DFT $\transform$ of the signal $\signal$, when the observed time-domain samples $\obs$ are corrupted by white Gaussian noise, i.e., 
\begin{equation}
\obs = \signal + \noise,
\end{equation} 
where $\noise \in \cnormal(0,\eye{\length}{\length})$. 

Further, we make the following assumptions on the signal:
\begin{itemize}
\item The number of non-zero DFT coefficients $\sparsity = O(\length^\sindex)$, where $ 0 \leq \sindex < 1$, i.e., sub-linear sparsity.
\item Let ${\magset} = \{\sqrt{\snr}/2 + i\sqrt{\snr}/\levels_1\}_{i=0}^{\levels_1}$ and $\phaseset= \{2\pi i/\levels_2\}_{i=0}^{\levels_2-1}$, where $\snr$ is a desired signal-to-noise ratio and $\levels_1,\levels_2$ are some constants.
\item We assume that all the non-zero DFT coefficients belong to the set $\constellation = \{\alpha e^{\imath\phi} \mid \alpha \in \magset, \phi \in \phaseset\}$ (see Remark~\ref{rem:constellation} for a further discussion on this constraint).
\item The support of the non-zero DFT coefficients is uniformly random in the set $\{0,1,\hdots,\length-1\}$.
\item The signal-to-noise ratio $\snr$, is defined as, $\snr = \expectation_{X[\ell] \neq 0}\{|X[\ell]|^2\}/\expectation\{||\noise||^2/\length\}$.
\end{itemize}

%*********************************************************************************%
% Main result
%*********************************************************************************%
\setcounter{subsubsection}{0}
\section{Main results}\label{sec:main}
The proposed R-FFAST architecture is targeted for applications, where a fixed sampling structure or front-end is used for all the input signals. It computes an $\length$-length $\sparsity$-sparse DFT $\transform$ using $O(\sparsity\log^3\length)$ noise-corrupted time-domain samples, in $O(\sparsity\log^4\length)$ complex operations, i.e., {\em sub-linear sample and time complexity}. A variant of the R-FFAST algorithm that allows for flexibility of choosing a different measurement matrix for every input signal is presented in \cite{li2014sub} and computes an $\length$-length $\sparsity$-sparse DFT $\transform$ using $O(\sparsity\log^{4/3}\length)$ samples and in $O(\sparsity\log^{7/3}\length)$ computations.

\begin{thm}\label{thm:main} For a finite but sufficiently high SNR, $0 < \sindex < 1$, and large enough length $\length$, the R-FFAST algorithm computes a $\sparsity$-sparse DFT $\transform \in \constellation^{\length}$, of an $\length$-length signal $\signal$, where $\sparsity = O(\length^{\sindex})$, from its noise-corrupted time-domain samples $\obs$, with the following properties:
\begin{itemize}
\item {\bf Sample complexity}: The algorithm needs $\samples = O(\sparsity\log^3\length)$ samples of $\obs$.
\item {\bf Computational complexity}: The R-FFAST algorithm computes DFT $\transform$ using $O(\sparsity\log^4\length)$ arithmetic operations.
\item {\bf Probability of success}: The algorithm perfectly recovers the DFT $\transform$ of the signal $\signal$, with probability at least $1 - O(1/\sparsity)$.
\end{itemize}
\begin{proof} Please see Appendix~\ref{app:main}.
\end{proof}
\end{thm}

\begin{remark}\label{rem:constellation} The R-FFAST algorithm reconstructs the $\sparsity$-sparse DFT $\transform$ perfectly, even from the noise-corrupted time-domain samples, with high probability, since in our signal model we assume that $\transform \in \constellation^{\length}$. The reconstruction algorithm is equally applicable to an arbitrary complex-valued non-zero DFT coefficients, as long as all the non-zero DFT coefficients respect the signal-to-noise-ratio as defined in Section~\ref{sec:model}. The proof technique for arbitrary complex-valued non-zero DFT coefficients becomes much more cumbersome, and we do not pursue it in this paper. However, in Section~\ref{sec:simulations} we provide an empirical evidence of the applicability of the R-FFAST algorithm to signals with arbitrary complex-valued DFT coefficients. The reconstruction is deemed successful if the support is recovered perfectly. We also note that the normalized $\ell_1$-error is small (see Section~\ref{sec:simulations} for more details) when the support is recovered successfully.
\end{remark}

%*********************************************************************************%
% Related work
%*********************************************************************************%
\setcounter{subsubsection}{0}
\section{Related work}\label{sec:related} 
The problem of computing a sparse discrete Fourier transform of a signal is related to the rich literature of frequency estimation \cite{prony1795essai, pisarenko1973retrieval, schmidt1986multiple, roy1989esprit} in statistical signal processing as well as compressive-sensing \cite{donoho2006compressed, candes2006near}. Most works in the frequency estimation literature use well studied subspace decomposition principles e.g., singular-value-decomposition, like MUSIC and ESPRIT \cite{prony1795essai, schmidt1986multiple, roy1989esprit}. These methods quickly become computationally infeasible as the problem dimensions $(\sparsity,\length)$ increase. In contrast, we take a different approach combining tools from coding theory, number theory, graph theory and statistical signal processing, to divide the original problem into many instances of simpler problems. The divide-and-conquer approach of R-FFAST alleviates the scaling issues in a much more graceful manner.

In compressive sensing, the bulk of the literature concentrates on random linear measurements, followed by either convex programming or greedy pursuit reconstruction algorithms \cite{candes2006near, candes2006robust, tropp2007signal}. A standard tool used for the analysis of the reconstruction algorithms is the {\em restricted isometry property} (RIP) \cite{candes2005decoding}. The RIP characterizes matrices which are nearly orthonormal or unitary, when operating on sparse vectors. Although random measurement matrices like Gaussian matrices exhibit the RIP with optimal scaling, they have limited use in practice, and are not applicable to our problem of computing a sparse DFT from time-domain samples. So far, to the best of our knowledge, the tightest characterization of the RIP, of a matrix consisting of random subset of rows of an $\length\times\length$ DFT matrix, provides a sub-optimal scaling, i.e., $O(\sparsity\log^4\length)$, of samples \cite{rauhut2012restricted}. In contrast, we show that by relaxing the worst case assumption on the input signal one can achieve a scaling of $O(k\log n)$ even for partial Fourier measurements. An alternative approach, in the context of sampling a continuous time signal with a finite rate of innovation is explored in \cite{vetterli2002sampling, dragotti2007sampling, blu2008sparse, mishali2010theory}.

At a higher level though, despite some key differences in our approach to the problem of computing a sparse DFT, our problem is indeed closely related to the spectral estimation and compressive sensing literature, and our approach is naturally inspired by this, and draws from the rich set of tools offered by this literature.   

A number of previous works \cite{GGI02, gilbert2008tutorial, hassanieh2012nearly, iwen2010combinatorial} have proposed a sub-linear time and sample complexity algorithms for computing a sparse DFT of a high-dimensional signal. Although sub-linear in theory, these algorithms require large number of samples in practice and are robust against limited noise models, e.g., bounded noise. {In \cite{ghazi2013sample}, the authors propose an algorithm for computing a $\sparsity$-sparse $2D$-DFT of an $\sqrt{\length}\times\sqrt{\length}$ signal, where $\sparsity = \Theta(\sqrt{\length})$. For this special case, the algorithm in \cite{ghazi2013sample} computes $2D$-DFT using $O(\sparsity\log\length)$ samples and $O(\sparsity\log^2\length)$ computations}. In contrast, the R-FFAST algorithm is applicable to $1D$ signals and for the entire sub-linear regime of sparsity, i.e., $\sparsity = O(\length^{\sindex}), 0< \sindex < 1$, when the time-domain samples are corrupted by Gaussian noise.
%*********************************************************************************%
% Sampling patterns and measurement matrix
%*********************************************************************************%
\setcounter{subsubsection}{0}
\section{R-FFAST architecture and algorithm exemplified}\label{sec:ExFFAST}
\begin{figure}[h]
 \begin{center}
\includegraphics[width = \linewidth]{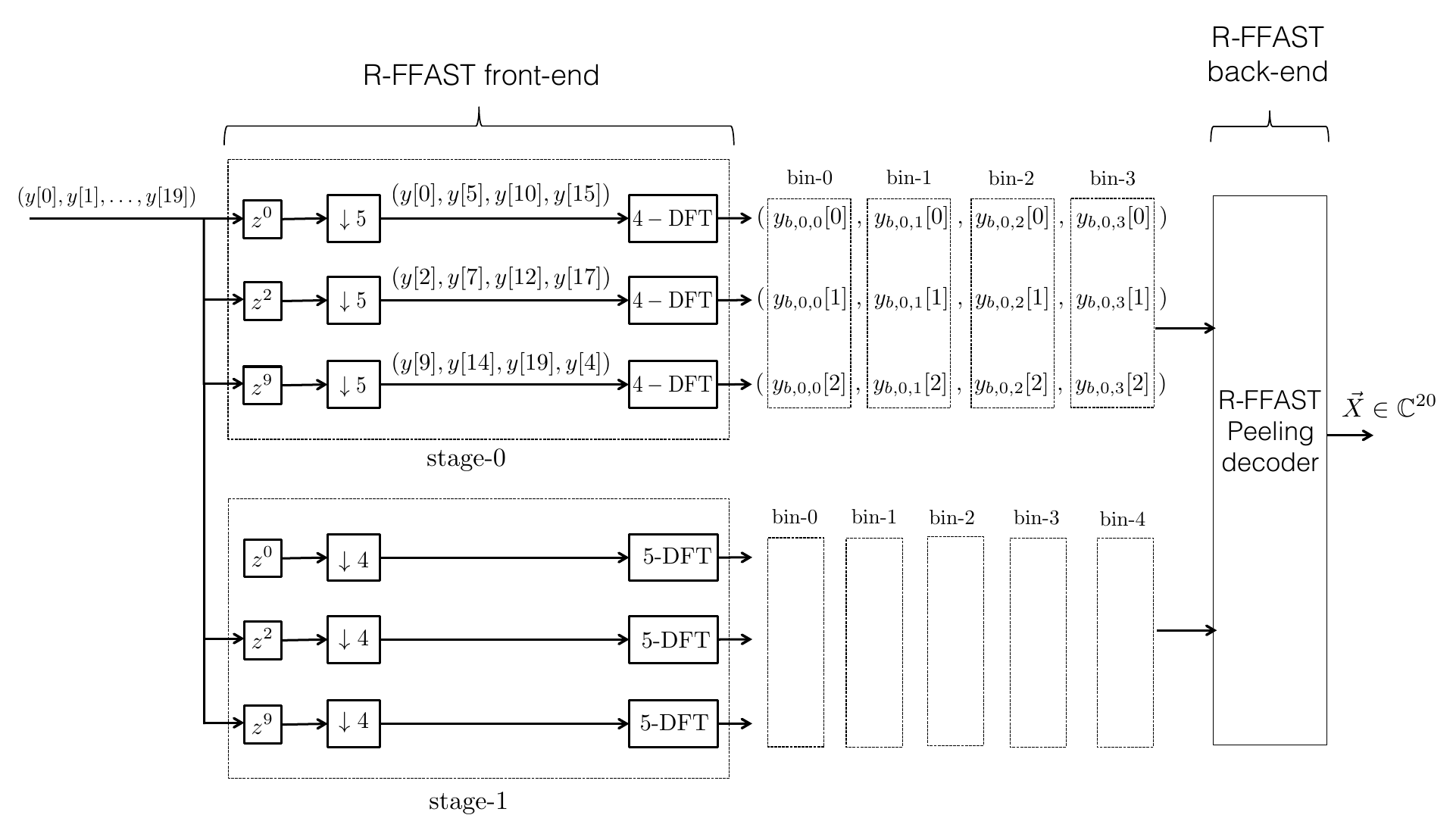}
\end{center}
\caption{An example $2$-stage R-FFAST architecture. The noise-corrupted time-domain samples $\obs = \signal + \noise$, of a $20$-point signal $\signal$ with a $5$ sparse DFT $\transform$, are processed using the R-FFAST front-end and the peeling back-end algorithm.}\vspace*{-0.1in}
\label{fig:ffastex}
\end{figure}

In this section, we describe the R-FFAST sub-sampling ``front-end" architecture as well as the associated ``back-end" peeling-decoder using a simple example. Later, in Section~\ref{sec:FFASTArch} and Section~\ref{sec:backEnd}, we provide a generic description of the front-end architecture and the back-end algorithm.

%\begin{example}
Consider an $\length = 20$ length input signal $\signal$, whose DFT $\transform$, is $\sparsity = 5$ sparse. Further, let the $5$ non-zero DFT coefficients of $\signal$ be $\transformc{1}$,  $\transformc{3}$, $\transformc{5}$, $\transformc{10}$ and $\transformc{15}$. Let $\obs = \signal + \noise$, be a $20$-length noise-corrupted time-domain signal, input to the R-FFAST architecture shown in Fig.~\ref{fig:ffastex}. In its most generic form, the R-FFAST sub-sampling `front-end' consists of $\stages$ ($\geq3$) stages, where $\stages$ is a non-decreasing function of the sparsity-index $\sindex$ (where $\sparsity = O(\length^{\sindex})$ ). Each stage further has $\delays$ number of subsampling paths or delay-chains with an identical sampling period, e.g., all the delay-chains in stage $0$ of the Fig.~\ref{fig:ffastex} have sampling period $5$. The input signal to the different delay-chains is circularly shifted by certain amount. For illustrative purposes, in Fig.~\ref{fig:ffastex}, we show the processing of noise-corrupted samples $\obs$ through a $\stages = 2$ stage R-FFAST architecture, where the sampling period of the two stages are $5$ and $4$  respectively. Each stage further has $\delays =3$ delay-chains. The input signal to each of the $\delays = 3$ delay-chains is circularly shifted by $0,2$ and $9$ respectively prior to the sub-sampling operation. The output of the R-FFAST front-end is obtained by computing the short DFTs of the sub-sampled data and further grouping them into `bins', as shown in Fig.~\ref{fig:ffastex}. The R-FFAST peeling-decoder then synthesizes the big $20$-point DFT $\transform$ from the output of the short DFTs that form the bin observations. 
\begin{figure}[h]
\begin{center}
\includegraphics[scale=.5]{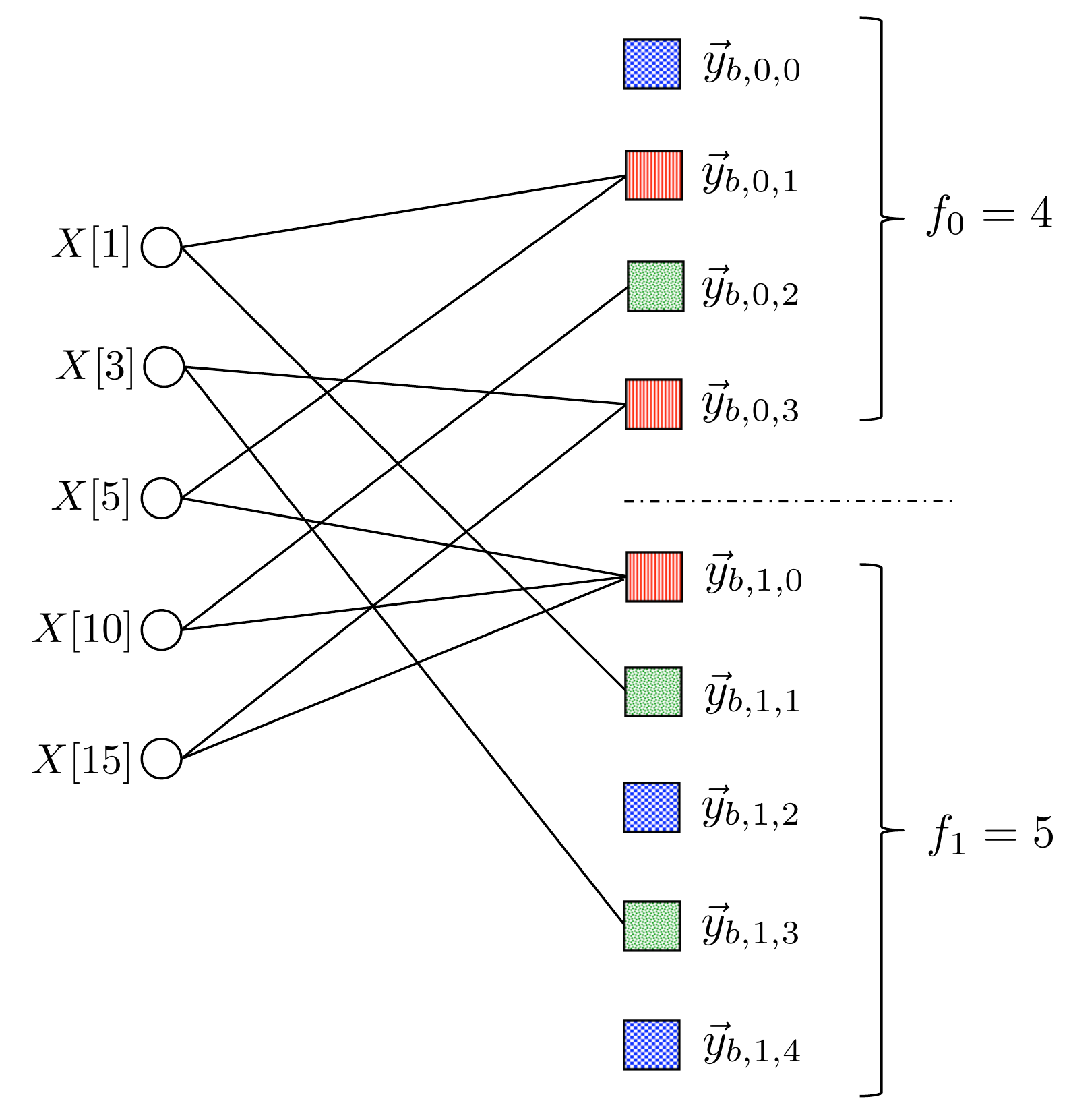}
\caption{A $2$-left regular degree bi-partite graph representing relation between the unknown non-zero DFT coefficients, of the $20$-length example signal $\signal$, and the bin observations obtained through the R-FFAST front-end architecture shown in Fig.~\ref{fig:ffastex}. Left nodes of the bi-partite graph represent the $5$ non-zero DFT coefficients of the input signal $\signal$, while the right nodes represent the ``bins" ( also sometimes referred in the sequel as ``check nodes") with vector observations. An edge connects a left node to a right node iff the corresponding non-zero DFT  coefficient contributes to the observation vector of that particular bin. The observation at each check node is a 3-dimensional complex-valued vector, e.g., $\binobsv{0}{0}$ is observation at bin $0$ of stage $0$.}
\label{fig:aliasinggraph}
\end{center}
\end{figure}
The relation between the output of the short DFTs, i.e., bin-observations, and the DFT coefficients of the input signal $\signal$, can be computed using preliminary signal processing identities. Let $\binobsv{i}{j}$ denote a $3$-dimensional observation vector of bin $j$ of stage $i$. Then, a graphical representation of this relation is shown using a bi-partite graph in Fig.~\ref{fig:aliasinggraph}. Left nodes of the bi-partite graph represent the $5$ non-zero DFT coefficients of the input signal $\signal$, while the right nodes represent the ``bins" with vector observations. We use $\factor{i}$ to denote the number of bins in stage-$i$, for $i=0,1$. An edge connects a left node to a right node iff the corresponding non-zero DFT coefficient contributes to the observation vector of that particular bin, e.g., after sub-sampling, the DFT coefficient $\transformc{10}$ contributes to the observation vector of bin $2$ of stage $0$ and bin $0$ of stage $1$. Thus, the problem of computing a $5$-sparse $20$-length DFT has been transformed into that of decoding the values and the support of the left nodes of the bi-partite graph in Fig.~\ref{fig:aliasinggraph}, i.e., sparse-graph decoding. Next, we classify the observation bins based on its edge degree in the bi-partite graph, which is then used to describe the FFAST peeling decoder.
\begin{itemize}
\item {\bf zero-ton}: A bin that has no contribution from any of the non-zero DFT coefficients of the signal, e.g., bin $0$ of stage $0$ in Fig.~\ref{fig:aliasinggraph}. 
\begin{eqnarray}
\binobsv{0}{0} =
\left(
\begin{array}{c}
 \nbinobsc{0}{0}[0] \\
 \nbinobsc{0}{0}[1]\\
 \nbinobsc{0}{0}[2]
 \end{array}
\right),
\end{eqnarray}
where $\nbinobsv{i}{j} \in \cnormal(0,\eye{3}{3})$ consists of the DFT coefficients of the samples of the noise vector $\noise$.
\item {\bf single-ton}: A bin that has contribution from exactly one non-zero DFT coefficient of the signal, e.g., bin $2$ of stage $0$. Using the signal processing properties, it is easy to verify that the observation vector of bin $2$ of stage $0$ is given as,
\begin{eqnarray}
\binobsv{0}{2} =
\left(
\begin{array}{c}
 1 \\
 e^{\imath2\pi20/20}\\
 e^{\imath2\pi90/20}
\end{array}
\right) \transformc{10} + \left(
\begin{array}{c}
 \nbinobsc{0}{2}[0] \\
 \nbinobsc{0}{2}[1]\\
 \nbinobsc{0}{2}[2]
 \end{array}
\right).
\end{eqnarray}
\item {\bf multi-ton}: A bin that has a contribution from more than one non-zero DFT coefficients of the signal, e.g., bin $1$ of stage $0$. The observation vector of bin $1$ of stage $0$ is,
\begin{eqnarray}\label{eq:binobs}
\binobsv{0}{1} &=&
\left(
\begin{array}{c}
 1\\
 e^{\imath2\pi2/20} \\
 e^{\imath2\pi9/20}
\end{array}
\right) \transformc{1} + \left(
\begin{array}{c}
 1\\
 e^{\imath2\pi10/20}\\
 e^{\imath2\pi45/20}
\end{array}
\right)\transformc{5} 
+ \left(
\begin{array}{c}
 \nbinobsc{0}{1}[0] \\
 \nbinobsc{0}{1}[1]\\
 \nbinobsc{0}{1}[2]
 \end{array}
\right).
\end{eqnarray}
\end{itemize}

%*********************************************************************************%
% Algorithm
%*********************************************************************************%
\begin{algorithm}[h]
\caption{R-FFAST Algorithm}
\label{alg:FFAST}
\begin{algorithmic}[1]
\STATE {\em Input:} The noise-corrupted bin observations $\binobsv{i}{j}$, obtained using the R-FFAST sub-sampling front-end, for each bin $j$ in stage $i$ for all $i,j$.
\algrule
\STATE {\em Output:}  An estimate $\transform$ of the $\sparsity$-sparse $\length$-point DFT.
\algrule
\STATE {\em R-FFAST Decoding:} Set the initial estimate of the $\length$-point DFT $\transform = 0$. Let $\iterations$ denote the number of iterations performed by the R-FFAST decoder.
\STATE Set the energy threshold $T = (1+\threshold)\delays$ for appropriately chosen $\threshold$ (see Appendix~\ref{app:main}).
\FOR {each iteration}
\FOR {each stage $i$}
\FOR {each bin $j$}
	\IF {$||\binobsv{i}{j}||^2 < T$}
		\STATE bin is a {\em zero-ton}. 
	\ELSE
		\STATE (singleton, $v_p$, $p$) = {\bf \emph{Singleton-Estimator}} $(\binobsv{i}{j})$.
	       \IF {singleton = `true'}
                        \STATE Peel-off: $\binobsv{s}{q} = \binobsv{s}{q} - v_p\svector{p}$, for all stages $s$ and bins $q \equiv p \text{ mod } \factor{s}$.
                      \STATE Set, $\transformc{p} = v_p$.
		\ELSE
			\STATE bin is a {\em multi-ton}.
		\ENDIF
	\ENDIF
\ENDFOR
\ENDFOR
\ENDFOR
\end{algorithmic}
\end{algorithm}

{\em \bf R-FFAST Iterative peeling-decoder}: 
The R-FFAST peeling decoder uses a function called ``singleton-estimator" (see Section~\ref{sec:backEnd} for details) that exploits the statistical properties of the bin observations to identify which bins are singletons and multitons with a high probability. In addition, if the bin is a singleton bin, it also determines the value and the support of the only non-zero DFT coefficient connected to this bin. The R-FFAST peeling decoder then repeats the following steps (see Algorithm~\ref{alg:FFAST} for the pseudocode):
\begin{enumerate}
\item Select all the edges in the graph with right degree 1 (edges connected to single-ton bins).
\item Remove these edges from the graph as well as the associated left and right nodes.
\item Remove all the other edges that were connected to the left nodes removed in step-2. When a neighboring edge of any right node is removed, its contribution is subtracted from that check node.
\end{enumerate}
Decoding is successful if, at the end, all the edges have been removed from the graph. It is easy to verify that the R-FFAST peeling-decoder operated on the example graph of Fig.~\ref{fig:aliasinggraph} results in successful decoding, with the coefficients being uncovered in the following possible order: $\transformc{10},\transformc{1},\transformc{3},\transformc{5},\transformc{15}$.
%\end{example}

The success of the R-FFAST peeling decoder depends on the following two crucial assumptions: 
\begin{enumerate}
\item The bipartite graph induced by the R-FFAST sub-sampling front-end is such that a singleton based iterative peeling decoder successfully uncovers all the variable nodes. 
\item The function ``singleton-estimator" reliably determines the identity of a bin using the bin-observations in the presence of observation noise.
\end{enumerate}

In \cite{pawar2013computing}, the authors have addressed the first assumption, of designing the FFAST sub-sampling front-end architecture, guided by the Chinese-remainder-theorem (CRT), such that the induced graphs (see Fig.~\ref{fig:aliasinggraph} for an example) belong to an ensemble of graphs that have a high probability of being decoded by a singleton based peeling decoder. For completeness, we provide a brief description of the FFAST sub-sampling front-end design in Section~\ref{sec:FFASTArch}, and refer interested readers to \cite{pawar2013computing} for a more thorough description. The second assumption of designing a low-complexity and noise-robust ``singleton-estimator" function is addressed in Section~\ref{sec:backEnd}. First, we review some results from the signal processing literature of {\em Frequency estimation} in the next section, that are key to designing the ``singleton-estimator" algorithm. 
\setcounter{subsubsection}{0}
\section{Frequency estimation: A single complex sinusoid} \label{sec:FreqEst} 
The problem of estimating the frequency and amplitude of a single complex sinusoid from noise corrupted time-domain samples has received much attention in signal processing literature. The optimal maximum likelihood estimator (MLE) is given by the location of the peak of a periodogram. However, the computation of periodogram is prohibitive even with an FFT implementation, and so simpler methods are desirable. In \cite{kay1989fast}, the author used a high SNR approximation from \cite{tretter1985estimating}, to propose an estimator that is computationally much simpler than the periodogram and yet attains the Cramer-Rao bound for moderately high SNRÕs. Next, we briefly review the high SNR approximation from \cite{tretter1985estimating}, along with the fast sinusoidal frequency estimation algorithm from \cite{kay1989fast}.

Consider $\samplesc$ samples of a single complex sinusoidal signal buried in white Gaussian noise, obtained at a periodic interval of period $T$,
\begin{eqnarray}\label{eq:FreqEst}
y(t) = A e^{j(\omega t + \phi)} + w(t), \ \ t = 0, T, 2T, \hdots, (\samplesc-1)T,
\end{eqnarray}
where the amplitude $A$, frequency $\omega$, and the phase $\phi$ are deterministic but unknown constants. The sampling period $T$ determines the maximum frequency $\omega_{\max} \in (0,2\pi/T)$, that can be estimated using these samples. Henceforth, for ease of notation WLOG we assume that $T=1$, i.e., $\omega_{\max} \in (0,2\pi)$. The sample noise $w(t)$ is assumed to be a zero-mean white complex Gaussian noise with variance $\sigma^2_w$. 

\subsubsection{High SNR approximation of the phase noise \cite{tretter1985estimating}} 
\begin{figure}[h]
 \begin{center}
\includegraphics[width = 0.6\linewidth]{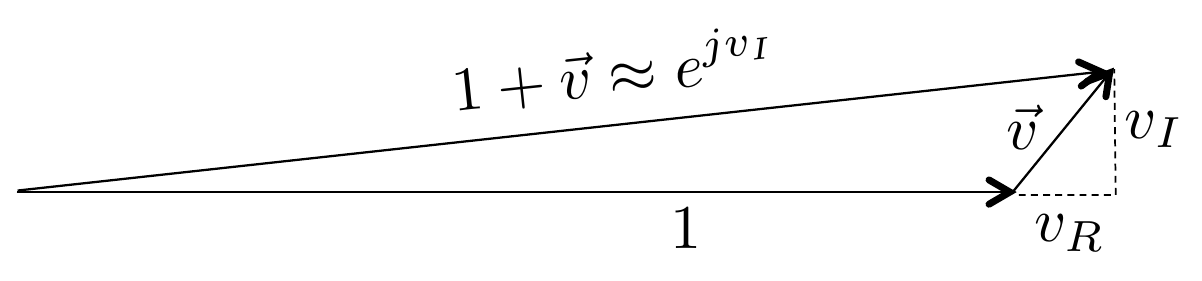}
\end{center}
\caption{Consider a complex number $1 + \vec{v}$, where $\vec{v} = v_R + j v_I$ and $|\vec{v}| << 1$. Then, a high SNR approximation of a complex number $1 + \vec{v}$  is given by $\exp(j v_I)$ \cite{tretter1985estimating}.}
\label{fig:highsnr}
\end{figure}
Let the signal-to-noise-ratio SNR $\snr$ be defined as $\snr = A^2/\sigma_w^2$. Consider the model of \eqref{eq:FreqEst},
\begin{eqnarray}
y(t) &=& A e^{j(\omega t + \phi)} + w(t) \nn\\
&=& (1 + v(t)) A e^{j(\omega t + \phi)}
\end{eqnarray}
where $v(t) = w(t)e^{-j(\omega t + \phi)}/A$, is a zero-mean complex normal random variable with variance $1/\snr$. For high SNR, i.e., $\snr >> 1$, we can approximate the complex number $(1 + v(t))$ as (also see Fig.~\ref{fig:highsnr}),
\begin{eqnarray}
(1 + v(t)) &=& \sqrt{(1 + v_R(t))^2 + v^2_I(t)}e^{j\tan^{-1}(v_I(t)/(v_R(t) + 1))}\nn\\
&\approx&1 e^{j\tan^{-1}(v_I(t)/(v_R(t) + 1))}\nn\\
&\approx&1 e^{j\tan^{-1}v_I(t)}\nn\\
&\approx&1 e^{jv_I(t)},
\end{eqnarray}
where the phase $v_I(t)$ is a zero mean Guassian random variable with variance $1/2\snr$.
The high SNR approximation model was first proposed in \cite{tretter1985estimating}. In \cite{kay1989fast}, the author empirically validated that the high SNR approximation holds for SNR's of order $5-7$dB.
Thus, under high SNR assumption the observations in \eqref{eq:FreqEst} can be written as,
\begin{eqnarray}\label{eq:highSNRapprox}
y(t) \approx A e^{j(\omega t + \phi + u(t))},
\end{eqnarray}
where $u(t)$ is zero-mean white Gaussian with variance $1/(2\snr)$.

\subsubsection{Frequency Estimator \cite{kay1989fast}}\label{sec:phasediff} Using the high SNR approximation model of \eqref{eq:highSNRapprox}, we have,
\begin{eqnarray}
\Delta(t) &=& \angle y(t+1) - \angle y(t), \ \ t = 0,1,\hdots,\samplesc-2 \nn \\
&=& \omega + u(t+1) - u(t) \nn \\
i.e., \ \vec{\Delta} &=& \vec{1} \ \omega + \vec{z},
\end{eqnarray}
where $\vec{z} \in \real^{\samplesc-1}$, is a zero-mean colored Gaussian noise. The author in \cite{kay1989fast} then computed the following MMSE rule to estimate the unknown frequency $\omega$,
\begin{equation}\label{eq:kayRule}
\hat{\omega} = \sum_{t = 0}^{\samplesc-2}\beta(t) \angle y(t+1)y^{\dagger}(t),
\end{equation}
where the weights $\beta(t)$ have closed form expression (see \cite{kay1989fast}) and are computed using the optimal whitening filter coefficients of an MMSE estimator. In \cite{kay1989fast}, the author shows that the frequency estimate of \eqref{eq:kayRule} is an unbiased estimator and is related to the true frequency $\omega$ as,
\begin{equation}\label{eq:kayEst}
\hat{\omega} = \left(\omega + \normal\left(0,\frac{6/\snr}{\samplesc(\samplesc^2-1)}\right)\right)_{2\pi},
\end{equation}
where $(\cdot)_{2\pi}$ is a modulo $2\pi$.
Equation~\eqref{eq:kayEst} essentially holds for any value of $T$, i.e., periodic sampling. 

In \cite{kay1989fast}, the author considered a problem where the unknown frequency $\omega$ is a real number in an appropriate range, e.g., $\omega \in (0,2\pi)$. Hence, the result in \eqref{eq:kayEst} is expressed in terms of an estimation error as a function of the number of measurements $N$ and the signal-to-noise ratio $\snr$. In contrast, in this paper we are interested in frequencies of the form $2\pi\ell/n$, for some unknown integer $\ell \in [0,\length-1]$. Hence, we use the Gaussian Q-function along with the estimate in \eqref{eq:kayEst} to get the following probabilistic proposition.
\begin{prop}\label{prop:clusterAccuracy} For a sufficiently high SNR and large $\length$, the frequency estimate $\hat{\omega}$ obtained using \eqref{eq:kayRule} satisfies,
\[
\pr(|\hat{\omega} - \omega| > \pi/\range) < 1/n^3,
\]
for any constant $\range$, where $\samplesc = O(\log^{1/3}\length)$.
\begin{proof} Please see Appendix~\ref{app:clusterAccuracy}
\end{proof}
\end{prop}
\setcounter{subsubsection}{0}
\section{R-FFAST front-end architecture}\label{sec:FFASTArch}
\begin{figure}[h!]
\begin{center}
\includegraphics[width = \linewidth]{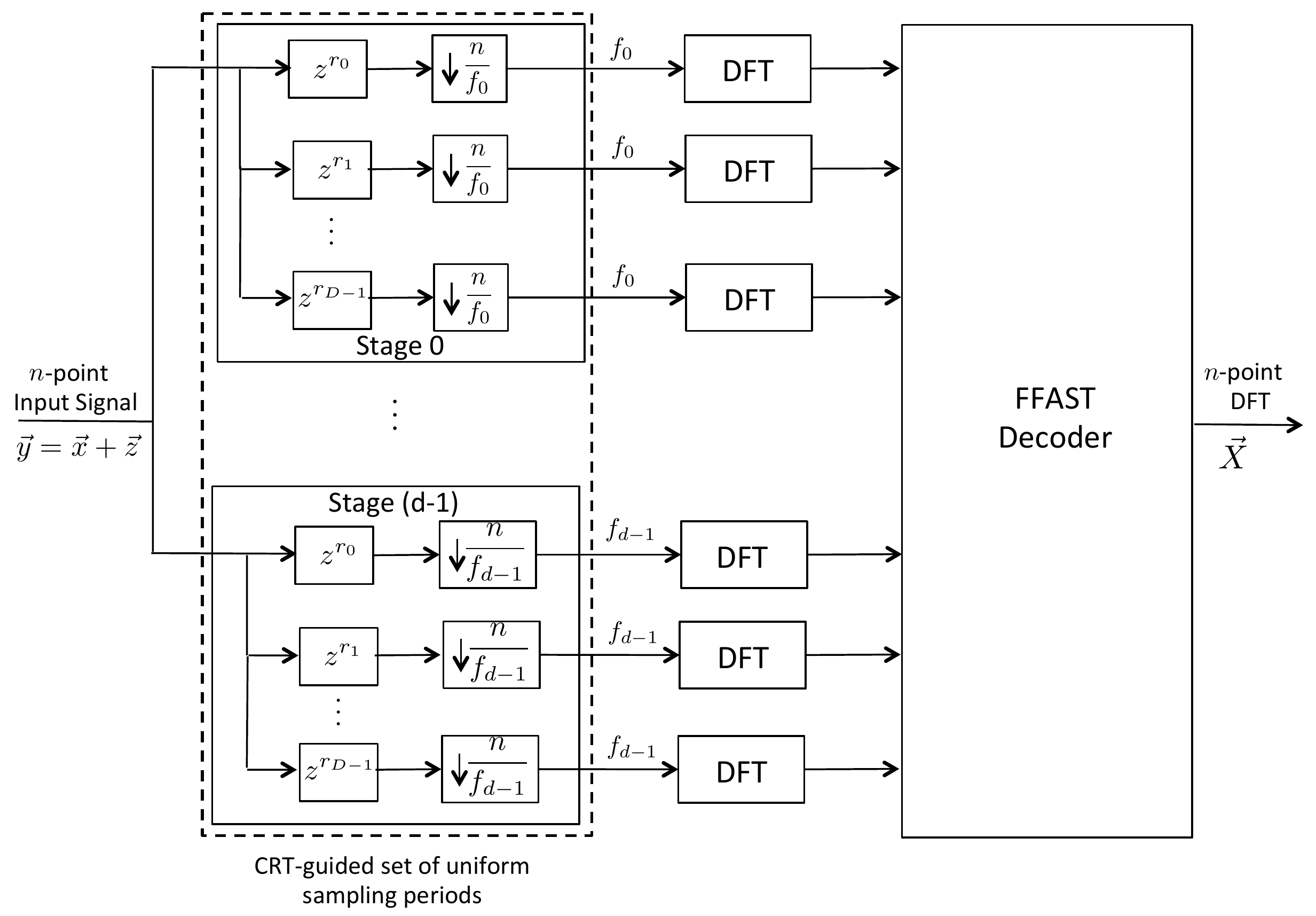}
\caption{A block diagram of the R-FFAST architecture for processing noise-corrupted samples $\obs = \signal + \noise$. The $\length$-point input signal $\obs$ is uniformly subsampled by a carefully chosen, guided by the Chinese-Remainder-Theorem, set of $\stages$ patterns. Each stage has $\delays$ delay-chains and a delay-chain in stage $i$ has $\factor{i}$ number of samples. The delay shift pattern $(r_0,r_1,\hdots,r_{\delays-1})$ is carefully chosen so as to induce bin-measurement matrices with ``good" mutual incoherence and RIP properties (as explained in Section~\ref{sec:delayCS}). Next, the (short) DFTs, of each of the delay-chain samples are computed using an efficient FFT algorithm of choice. The big $\length$-point DFT $\transform$ is then synthesized from the smaller DFTs using the peeling-like R-FFAST decoder.}
\label{fig:conceptual}
\end{center}
\end{figure}

In its most generic form the R-FFAST front-end architecture consists of $\stages \geq 3$ stages and $\delays$ delay-chains per stage as shown in Fig.~\ref{fig:conceptual}. The $i^{th}$ delay-chain in stage-$j$, circularly shifts the input signal by $r_i$ and then sub-samples by a sampling period of $\length/\factor{j}$, to get $\factor{j}$ output samples. Using identical circular shifts $r_i$ in the $i^{th}$ delay-chain of all the stages is not a fundamental requirement of our architecture, but a convenience that is sufficient, and makes the R-FFAST framework implementation friendly. Next, the (short) DFTs of the output samples of each delay-chain are computed using an efficient FFT algorithm of choice. The big $\length$-point DFT $\transform$ is then synthesized from the smaller DFTs using the peeling-like R-FFAST decoder.

\subsection{Number of stages and sampling periods}\label{sec:nbins}
The number of stages $\stages$ and the sub-sampling factor's $\length/\factor{i}$, for $i=0,\hdots,\stages-1$, are chosen based on the sparsity-index $\sindex$, where $\sparsity = O(\length^{\sindex}), 0 \leq \sindex < 1$. Based on the qualitative nature of the design parameters, of the R-FFAST front-end architecture, we identify two regimes of sub-linear sparsity; 1) {\em very-sparse} regime, where $0 \leq \sindex \leq 1/3$, and 2) {\em less-sparse} regime, where $1/3 < \sindex < 1$. The role of the number of stages and the subsampling factors used in each stage of R-FFAST front-end is to guarantee that the frequency domain aliasing sparse bi-partite graph (see Fig.~\ref{fig:aliasinggraph} for an example) is successfully decoded by a singleton based peeling-decoder described in Section~\ref{sec:ExFFAST}, with a high probability. The choice of number of stages and the subsampling factors in R-FFAST front-end architecture is identical to that of the FFAST front-end architecture \cite{pawar2013computing}. Here, we provide a brief description about the choice of these parameters and refer interested reader's to \cite{pawar2013computing} for more details.
\begin{itemize}
\item Very-sparse regime: For the very-sparse regime we always use $\stages = 3$ stages in the R-FFAST front-end. The sampling periods $\length/\factor{i}$, for stages $i=0,1,2$, are chosen such that the integers $\factor{i} = O(\sparsity) + O(1)$, i.e., are approximately equal to $\sparsity$, and are relatively co-prime factors of $\length$. For example, $\length = 100*101*103*99$, $\sparsity = 100$, $\factor{0} = 100, \factor{1} = 101$ and $\factor{2} = 103$, achieves an operating point of $\sindex = 1/4$. 
\item Less-sparse regime: For the less-sparse regime the number of stages $\stages$ is a monotonic non-decreasing function of the sparsity index $1/3 < \sindex < 1$. In particular, an operating point of fractional sparsity index $\sindex$ can be achieved as follows. Let the number of stages $\stages = 1/(1-\sindex)$. Consider a set of $\stages$ relative co-prime integers $\primef{i} = O(\sparsity^{1/(\stages-1)})$ such that $\length = \prod_{i=0}^{\stages-1}\primef{i} = O(\sparsity^{1/\sindex})$. Then, the subsampling periods $\length/\factor{i}$ are chosen such that $\factor{i} = \prod_{j=i}^{i+\stages-2}\primef{(j)_{\stages}}$, where $(j)_{\stages} = j \mod \stages$. For example, $\length = 10*11*13$, $\sparsity = 100$ and $\factor{0} = 10*11, \factor{1} = 11*13, \factor{2} = 13*10$, achieves $\sindex = 2/3$.
\end{itemize}

\subsection{Delay chains and the bin-measurement matrices}\label{sec:delayCS}
In the FFAST front-end architecture proposed in \cite{pawar2013computing}, for the noiseless observations, we have shown that $\delays = 2$ delays are sufficient to detect the identify of a bin from its observations. On the contrary, in the presence of additive noise, as will be shown in the sequel, each stage of the R-FFAST architecture needs many more delay-paths to average out the effect of noise. In order to understand the role of the number and the structure of the delay-chains, used in the R-FFAST front-end architecture, we take a detailed look at the bin-observations of the example of Section~\ref{sec:ExFFAST}. 

Consider the observation vector $\binobsv{0}{1}$, shown in \eqref{eq:binobs}, of the bin $1$ of stage $0$,
\begin{eqnarray*}
\binobsv{0}{1} &=&
\left(
\begin{array}{c}
 1\\
 e^{\imath2\pi2/20} \\
 e^{\imath2\pi9/20}
\end{array}
\right) \transformc{1} + \left(
\begin{array}{c}
 1\\
 e^{\imath2\pi10/20}\\
 e^{\imath2\pi45/20}
\end{array}
\right)\transformc{5} 
+ \left(
\begin{array}{c}
 \nbinobsc{0}{1}[0] \\
 \nbinobsc{0}{1}[1]\\
 \nbinobsc{0}{1}[2]
 \end{array}
\right),\\
&=& \left(
\begin{array}{ccccc}
\svector{1} & \svector{5} &  \svector{9} & \svector{13} & \svector{17}
\end{array}
\right)  \left(
\begin{array}{c}
 \transformc{1}  \\
 \transformc{5} \\
 \transformc{9} \\
 \transformc{13}\\
 \transformc{17}   
 \end{array}
\right) + \left(
\begin{array}{c}
 \nbinobsc{0}{1}[0] \\
 \nbinobsc{0}{1}[1]\\
 \nbinobsc{0}{1}[2]
 \end{array}
\right),\\
&=& \mmatrix_{0,1}\transform + \nbinobsv{0}{1},
\end{eqnarray*}
where the $\mmatrix_{0,1}$ is referred to as a ``bin-measurement" matrix of bin $1$ of stage $0$, and its $\ell^{th}$ column is given by \begin{equation*}
\svector{\ell} = \left\{\begin{array}{cc}
   \left(
\begin{array}{c}
  1\\
  e^{\imath2\pi2\ell/20}\\
  e^{\imath2\pi9\ell/20}
\end{array}
\right)& \text{ if } \ell \equiv 1 \text{ mod } 4\\
  \zvector & \text{otherwise},
\end{array}\right.
\end{equation*}
for $\ell = 0,1,\hdots,19$. In the sequel, we refer to $\svector{\ell},  \ \ell = 0,1\hdots, 19$, as a {\em steering-vector} corresponding to the frequency $2\pi\ell/20$, sampled at times $0,2$ and $9$.  

More generally, the observation vector $\binobsv{i}{j}$, of bin $j$ of stage $i$ of a R-FFAST front-end architecture with $\stages$ stages and $\delays$ delay-chains per stage, is given by,
\begin{eqnarray}\label{eq:subproblem}
\binobsv{i}{j} &=& \mmatrix_{i,j}\transform + \nbinobsv{i}{j}, \ \ 0 \leq i < \stages, \ 0 \leq j < \factor{i},
\end{eqnarray}
where $\factor{i}$ is the number of samples per delay-chain in stage $i$, $\nbinobsv{i}{j}\sim \cnormal({0,\eye{\delays}{\delays}})$ and $\mmatrix_{i,j}$ is the {\em bin-measurement} matrix. The bin-measurement matrix $\mmatrix_{i,j}$ is a function of the sampling period, number of delay-chains as well as the circular shifts $\{r_s\}_{s=0}^{\delays-1}$ used in stage $i$ of the R-FFAST sub-sampling front-end and its $\ell^{th}$ column is given by
\begin{equation}\label{eq:steeringVec}
\svector{\ell} = \left\{\begin{array}{cc}
  \left(
\begin{array}{c}
  e^{-\frac{\imath2\pi\ell r_0}{\length}},
  e^{-\frac{\imath2\pi\ell r_1}{\length}},
  \hdots,
  e^{-\frac{\imath2\pi\ell r_{\delays-1}}{\length}}
\end{array}
\right)^{\dagger} & \text{ if } \ell \equiv j \text{ mod } \factor{i}\\
  \zvector & \text{otherwise},
\end{array}\right.
\end{equation}
Note that, the $\ell^{th}$ column of the bin-measurement matrix is either zero or a sinusoidal frequency $2\pi\ell/\length$ sampled at times $\{r_s\}_{s=0}^{s=\delays-1}$.

In \eqref{eq:subproblem}, both the measurement matrix and the DFT vector are sparse. The singleton-estimator function attempts to determine if the bin observation $\binobsv{i}{j}$,  corresponds to a $0$-sparse (zeroton) or $1$-sparse (singleton) or $L$-sparse ($L \geq 2$ multi-ton) signal. In the case when the bin observation corresponds to a singleton bin, the function singleton estimator also tries to recover the $1$-sparse signal. Thus, the bin estimation problem is somewhat similar to {\em compressed sensing} problem. In the compressed sensing literature the {\em mutual-incoherence} and the {\em Restricted-isometry-property} (RIP) \cite{candes2005decoding}, of the measurement matrix, are widely studied and well understood to play an important role in stable recovery of a sparse vector from linear measurements in the presence of observation noise. Although, these techniques are not necessary for our bin-estimation problem, as we will see in the following, they are certainly sufficient. Next, we define the mutual-incoherence and the RIP of the bin-measurement matrices. Later, we use these properties to prove a stable recovery of the proposed R-FFAST peeling-style iterative recovery algorithm. 

\begin{definition}\label{def:coherence} The mutual incoherence $\maxcoh(\mmatrix)$ of a measurement matrix $\mmatrix$ is defined as 
\begin{equation}\label{eq:coherence}
\maxcoh(\mmatrix) \triangleq \max_{\forall p\neq q} \frac{|\vec{a}(p)^{\dagger}\vec{a}(q)|}{||\vec{a}(p)||\cdot||\vec{a}(q)||},
\end{equation}
where $\vec{a}(p)$ is the $p^{th}$ column of the matrix $\mmatrix$.
\end{definition}
The mutual-incoherence property of the measurement matrix indicates the level of correlation between the distinct columns of the measurement matrix. Smaller value of $\maxcoh(\mmatrix)$ implies more stable recovery, e.g., $\maxcoh(\mmatrix) = 0$ for an orthogonal measurement matrix $\mmatrix$.

\begin{definition}\label{def:RIP} The restricted-isometry constant $\ripc_s(\mmatrix)$ of a measurement matrix $\mmatrix$, with unit norm columns, is defined to be the smallest positive number that satisfies
\begin{equation}\label{eq:RIP}
(1 - \ripc_s(\mmatrix)) ||\transform||^2 \leq ||\mmatrix\transform||^2 \leq (1 + \ripc_s(\mmatrix)) ||\transform||^2,
\end{equation} 
for all $\transform$ such that $\zeronorm{\transform} \leq s$.
\end{definition}
The RIP characterizes the norm-preserving capability of the measurement matrix  when operating on sparse vectors. If the measurement matrix $\mmatrix$ has a good RIP constant (small value of $\ripc_{2s}(\mmatrix)$) for all $2s$-sparse vectors, then a stable recovery can be performed for any $s$-sparse input vector \cite{candes2006stable}. Since, a small value of $\ripc_{2s}(\mmatrix)$ implies that $||\mmatrix(\transform_1 - \transform_2)||^2$ is bounded away from zero for any two distinct, $\transform_1\neq \transform_2$, $s$-sparse vectors.
\subsubsection{Randomness and reliability} As pointed out in earlier discussion, a smaller value of mutual incoherence corresponds to more reliability or stable recovery. If the circular shifts $\{r_s\}_{s=0}^{\delays-1}$ used in the $\delays$ delay chains of the R-FFAST front-end architecture are chosen uniformly at random between $0$ and $\length-1$, then,  
\begin{lemma}\label{lem:maxcoherence} The mutual incoherence $\maxcoh(\mmatrix_{i,j})$ of a bin-measurement matrix $\mmatrix_{i,j}$ is upper bounded by
\begin{equation}\label{eq:maxcoherence}
\maxcoh(\mmatrix_{i,j}) < 2\sqrt{\log(5\length)/\delays} \ , \ \forall \ i,j
\end{equation}
with probability at least $0.2$.
\begin{proof}
Please see Appendix~\ref{app:maxcoherence}.
\end{proof}
\end{lemma}
Also, it is easy, i.e., $O(\length\delays)$ complexity, to verify if a given choice of $r_0,\hdots,r_{\delays-1}$ satisfies \eqref{eq:maxcoherence} or not. Hence, using offline processing one can choose a pattern $(r_0,\hdots,r_{\delays-1})$, such that deterministically the bound in \eqref{eq:maxcoherence} is satisfied. In the following Lemma, we use the Gershgorin circle theorem \cite{gershgorin1931uber} to obtain a bound on the RIP condition of a matrix as a function of its mutual incoherence.

\begin{lemma}\label{lem:RIP} A matrix $\mmatrix$, with unit norm columns and mutual incoherence $\maxcoh(\mmatrix)$, satisfies the following RIP condition for all $\transform$ that have $\zeronorm{\transform} \leq s$,
\begin{equation}\label{eq:RIP}
\left(1 - \maxcoh(\mmatrix)(s-1)\right)^{+} ||\transform||^2\leq ||\mmatrix\transform||^2 \leq \left(1 + \maxcoh(\mmatrix)(s-1)\right)||\transform||^2 \nn
\end{equation}
\begin{proof}
Please see Appendix~\ref{app:RIP}.
\end{proof}
\end{lemma}

\subsubsection{Sinusoidal structured bin-measurement matrix for speedy recovery} In Section~\ref{sec:delayCS}, we have shown that the columns of a bin-measurement matrix $\mmatrix_{i,j} \in \complex^{\delays\times \length}$ are determined by circular shifts used in each of the delay-chains in the R-FFAST front-end architecture. In particular, as shown in \eqref{eq:steeringVec}, the $\ell^{th}$ column of a bin-measurement matrix is either zero or a sinusoidal signal with frequency $2\pi\ell/\length$ sampled at random times $\{r_s\}_{s=0}^{s=\delays-1}$ corresponding to the circular shifts used in the R-FFAST front-end. Although a uniformly random choice of circular shifts endows the bin-measurement matrix with good coherence properties, as shown in Lemma~\ref{lem:maxcoherence}, the lack of structure makes the decoding process, i.e., bin-processing algorithm, painfully slow. In contrast, if the columns of the bin-measurement matrix consist of {\em equispaced samples} of sinusoidal frequency $2\pi\ell/\length$, then as seen in Section~\ref{sec:FreqEst}, the sinusoidal signal can be detected using only linear operations, but the mutual incoherence properties are not sufficient enough for reliable decoding. 

In this section, we propose a choice of the circular shifts, used in the R-FFAST front-end architecture, that is a combination of the {\em randomness} and {\em structured sampling} that enables us to achieve {\em the best of both worlds}. We split the total number of delay-chains $\delays$ into $\clusters$ clusters, with each cluster further consisting of $\samplesc$ delay-chains, i.e., $\delays= \samplesc\clusters$. The first delay-chain of each cluster circular shifts the input signal by $r_i, i = 0,\hdots,\clusters-1$, a uniformly random number between $0$ and $\length-1$. The $j^{th}$ delay-chain of the $i^{th}$ cluster circularly shifts the input signal by $r_i + (j-1)2^{i}$, i.e., equisapced samples with spacing of $2^i$ starting from $r_i$. The $\ell^{th}$ column of the resultant bin-measurement matrix is given by, 
\begin{eqnarray}\label{eq:msvector}
\svector{\ell} &=& 
\left(
\begin{array}{c}
  e^{\imath 2\pi\ell (r_0+0)/n}\\
  e^{\imath 2\pi\ell (r_0+1)/n}\\
  \vdots\\
  e^{\imath 2\pi\ell (r_0+(\samplesc-1))/n}\\
  e^{\imath 2\pi\ell (r_1+0)/n}\\
  e^{\imath 2\pi\ell (r_1+2)/n}\\
  \vdots\\
  e^{\imath 2\pi\ell (r_1+2(\samplesc-1))/n}\\
  \vdots\\
  e^{\imath 2\pi\ell (r_{\clusters-1}+2^{\clusters-1}(\samplesc-1))/n}
\end{array}
\right).
\end{eqnarray}
In next section, we describe a bin-processing algorithm that exploits the equispaced structure, of the intra-cluster samples, to estimate the support of a singleton bin in linear time of the number of samples, i.e., $O(\delays)$.

\setcounter{subsubsection}{0}
\section{R-FFAST back-end peeling algorithm}\label{sec:backEnd}
The R-FFAST back-end peeling decoder recovers the big $\length$-point DFT $\transform$ from the smaller DFTs output by the R-FFAST front-end, i.e., bin-observations, using an onion peeling-like algorithm as described in Algorithm~\ref{alg:FFAST}. One of the crucial sub-routine used by the R-FFAST back-end peeling algorithm is ``singleton-estimator". The function singleton estimator exploits the statistical properties of the bin observations to identify which bins singletons and multitons with a high probability. In addition, if the bin is a singleton bin, it also determines the value and the support of the only non-zero DFT coefficient connected to this bin. In Section~\ref{sec:ExFFAST}, we explained the operation of the R-FFAST back-end peeling algorithm using a simple example. In this section, we focus on the description of the ``singleton-estimation" function. 

The singleton-estimator function estimates a single column of the bin-measurement matrix (or equivalently the frequency $\omega = 2\pi\ell/\length$) that best describes (in squared error sense) the bin observations. It does so using a two step procedure. First, the frequency estimation algorithm in \cite{kay1989fast} (i.e., estimator in \eqref{eq:kayRule}) is used to process the observations corresponding to each cluster, to obtain $\clusters$ estimates $\omega_i$ for $i=0,\hdots,\clusters-1$. Then, these estimates are fused together to obtain the final frequency estimate $\hat{\omega}$ that corresponds to the support of the potential singleton bin. If the estimated column does not justify the actual bin observations to a satisfactory level (based on an appropriately chosen threshold) then the processed bin is declared to be a multi-ton bin.

\begin{algorithm}[t]
\caption{singleton-estimator}
\label{alg:binprocess}
\begin{algorithmic}[1]
\STATE {\em Inputs:} 
\begin{itemize}
\item A noise-corrupted bin observations $\obs_b \in \complex^{\samplesc\clusters}$, consisting of $\clusters$ clusters each with $\samplesc$ number of measurements.
\item MMSE filter coefficients $\{\beta(0),\hdots,\beta(\samplesc-2)\}$.
\end{itemize}
\algrule
\STATE {\em Outputs:} 1) A boolean flag `singleton', 2) If singleton, then estimate of the value $v_p$ of the non-zero DFT coefficient and the position (or support) $p$ of the non-zero DFT coefficient.
\algrule
\STATE Set $\omega_{prev} = 0$.
\STATE Set the singleton = `false'.
\STATE Set $\delays = \samplesc\clusters$.
\FOR {each cluster $i=0,\hdots, \clusters-1$}
	\STATE $\omega_{new} = \frac{1}{2^i}\sum_{t = \samplesc i}^{\samplesc(i+1)-2}\beta(t-\samplesc i) 
		\angle y_b(t+1)y_b^{\dagger}(t)$.
	\STATE $\delta_1 = \lceil \frac{\omega_{prev}}{2\pi/2^i}\rceil (2\pi/2^i) + 
		\omega_{new} - \omega_{prev}$.
	\STATE $\delta_2 = \lfloor \frac{\omega_{prev}}{2\pi/2^{i}}\rfloor (2\pi/2^i) + 	
		\omega_{new} - \omega_{prev}$.
	\IF {$|\delta_1| < |\delta_2|$}
		\STATE $\omega_{prev} = \delta_1 + \omega_{prev}$.
	\ELSE
		\STATE $\omega_{prev} = \delta_2 + \omega_{prev}$.
	\ENDIF
\ENDFOR
\STATE Set the support estimate $q = round(\omega_{prev}n/2\pi)$.
\STATE Set the energy threshold $T = (1+\threshold)\delays$ for an appropriately chosen $\threshold$ (see Appendix~\ref{app:main}).
\STATE $v_q = \svector{q}^{\dagger}\obs_b/\delays$.
\IF {$||\obs - v_q\svector{q}||^2 < T$}
	\STATE singleton = `true'.
	\STATE $p = q$ and $v_p = v_q$.
\ENDIF
\end{algorithmic}
\end{algorithm}

\subsubsection{Processing observations in a cluster} Consider processing the observations of cluster $i$.
\begin{eqnarray}\label{eq:clusterobs}
\Delta_i(t) &=& \angle y(t+1) - \angle y(t), \ \ t = \samplesc i,\samplesc i+1,\hdots,\samplesc(i+1)-2 \nn \\
&=& 2^i\omega + u(t+1) - u(t) \nn \\
\vec{\Delta}_i &=& \vec{1} \ 2^i\omega + \vec{z},
\end{eqnarray}
where we have used the high SNR approximation model to get $\vec{z} \in \real^{\samplesc-1}$ as a zero-mean colored Gaussian noise. Further, applying the MMSE rule of \eqref{eq:kayRule} we get an estimate $\omega_i$ of $2^i\omega$ as follows,
\begin{eqnarray}
\omega_i = \left(2^i\omega + \normal\left(0,\frac{6/\snr}{\samplesc(\samplesc^2-1)}\right) \right)_{2\pi}
\end{eqnarray}

Thus, each cluster of observations provides an estimate (up to modulo $2\pi$) of a multiple of the true frequency $\omega$. Next, we show how these different estimates can be used to successively refine the search space of the true frequency $\omega$.

\subsubsection{Successive refinement}\label{sec:refine}
Let $\Omega_i = (\omega_i-\pi/\range, \omega_i+\pi/\range)$ be a range of frequencies around the estimate $\omega_i$ that contains the true frequency $2^{i}\omega$ with high a probability (as shown in proposition~\ref{prop:clusterAccuracy}), obtained by processing observations of cluster-$i$. For a continuous set, with a slight abuse of
notation we use $|\cdot|$ to denote the length of the interval, e.g., $|\Omega_0| = 2\pi/\range$. The operation of multiplying a frequency by $2$, maps two frequencies $\theta$ and $\theta+\pi$ to the same frequency, i.e., $(2\theta)_{2\pi} = (2(\theta + \pi))_{2\pi}$. However, since $|\Omega_0| = 2\pi/\range$, for a sufficiently large constant $\range$, $|\Omega_0 \cap \Omega_1/2| \leq 2\pi/(2\range)$. Similarly, the intersection of estimates from $\clusters$ consecutive clusters is,
\begin{eqnarray}\label{eq:cintersect}
|\cap_{i=0}^{\clusters-1}\Omega_i/2^{i}| \leq \frac{2\pi}{2^{\clusters-1}\range}.
\end{eqnarray}
Further, for a singleton bin, we know that the frequency corresponding to the location of the non-zero DFT coefficient is of the form $2\pi\ell/n$ for some integer $\ell \in \{0,\hdots,n-1\}$. Hence, if the number of clusters $\clusters = O(\log\length)$ and the number of samples per cluster\footnote{In construction of the measurement matrix \eqref{eq:msvector}, as well as in algorithm~\ref{alg:binprocess}, we assumed that the signal length $\length$ is not an even number. If $\length$ is an even number, then one can use some other prime, not a factor of $\length$, in  \eqref{eq:msvector} and in algorithm~\ref{alg:binprocess}. For example, if $\length$ is even but does not have $3$ as a factor, then the samples in $i^{th}$ cluster are equispaced with difference being $3^{i}$, instead of $2^i$ as suggested in \eqref{eq:msvector}.} $\samplesc= O(\log^{1/3}\length)$, then the singleton-estimator algorithm correctly identifies the unknown frequency with a high probability (see proposition~\ref{prop:DAccuracy}). The pseudo code of the algorithm is provided in Algorithm~\ref{alg:binprocess}.

\begin{prop}\label{prop:DAccuracy} For a sufficiently high SNR and large $\length$, the singleton-estimator algorithm with $\clusters = O(\log\length)$ clusters and $\samplesc = O(\log^{1/3\length})$ samples per cluster, i.e., $\delays = O(\log^{4/3}\length)$ samples, correctly identifies the support of the non-zero DFT coefficient contributing to the singleton bin with probability at least $1 - O(1/\length^2)$. The algorithm uses no more than $O(\log^{4/3}\length)$ complex operations.
\begin{proof} Please see Appendix~\ref{app:DAccuracy}
\end{proof}
\end{prop}
%*********************************************************************************%
% Simulations
%*********************************************************************************%
\section{Simulation Results}\label{sec:simulations}
In this section, we first evaluate the performance of the R-FFAST algorithm on synthetic data and contrast the results with the theoretical claims of Theorem~\ref{thm:main}. The synthetic data used for evaluation is beyond the theoretical claims of Theorem~\ref{thm:main}. In particular, the DFT coefficients of the used synthetic signal have arbitrary phase, instead of a finite constellation as assumed in Section~\ref{sec:model}. For arbitrary complex-valued DFT coefficients one cannot expect to perfectly reconstruct the DFT $\transform$ from noise-corrupted time-domain samples. A reconstruction is deemed successful, if the support of the non-zero DFT coefficients is recovered perfectly. Additionally, we observe that the normalized $\ell_1$-error is small compared to $1$ when support recovery is successful. 

In addition to evaluating the performance of the R-FFAST algorithm on synthetic signals, in Section~\ref{sec:sim_brain}, we also provide details of the application of a $2D$ version of the R-FFAST algorithm to acquire the Magnetic Resonance Image (MRI). Application of the R-FFAST algorithm to acquire MRI is not meant to compete with the state-of-the-art techniques in the MRI literature, but rather to demonstrate the feasibility of our proposed approach as a promising direction for future research.  More significantly, this experiment demonstrates that while the theory proposed in this paper applies to signals with a uniformly-random support  model for the dominant sparse DFT coefficients,  in practice, our algorithm works even for the non-uniform (or clustered) support setting as is typical of MR images.

\subsection{R-FFAST performance evaluation for synthetic signals}

\begin{figure}[t]
\centering     
\subfigure[Sample complexity of the R-FFAST algorithm as a function of signal length $\length$.]{\label{fig:svnplot}\includegraphics[width=.4\linewidth]{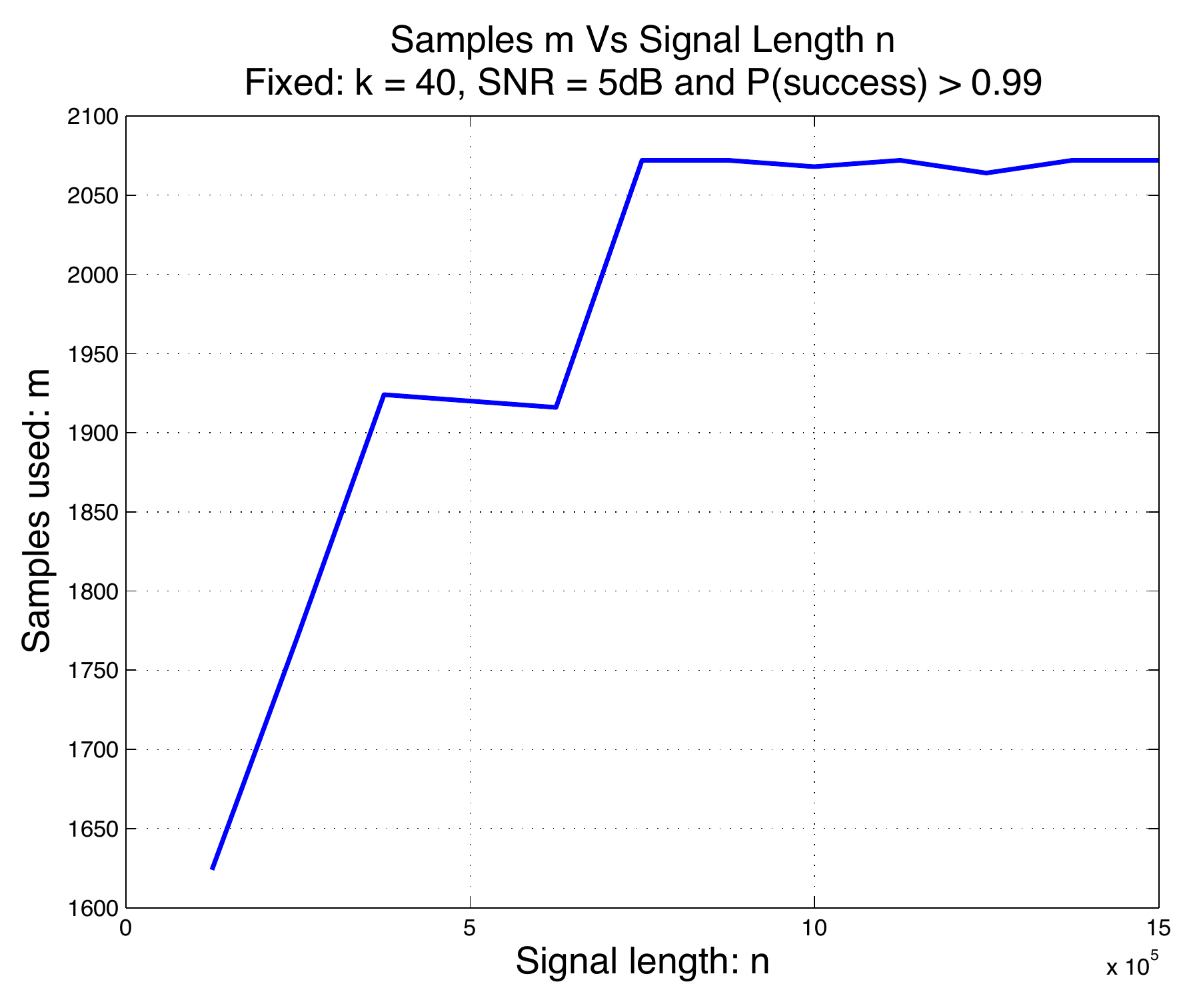}}
\subfigure[Time complexity of the R-FFAST algorithm as a function of signal length $\length$.]{\label{fig:tvnplot}\includegraphics[width=.4\linewidth]{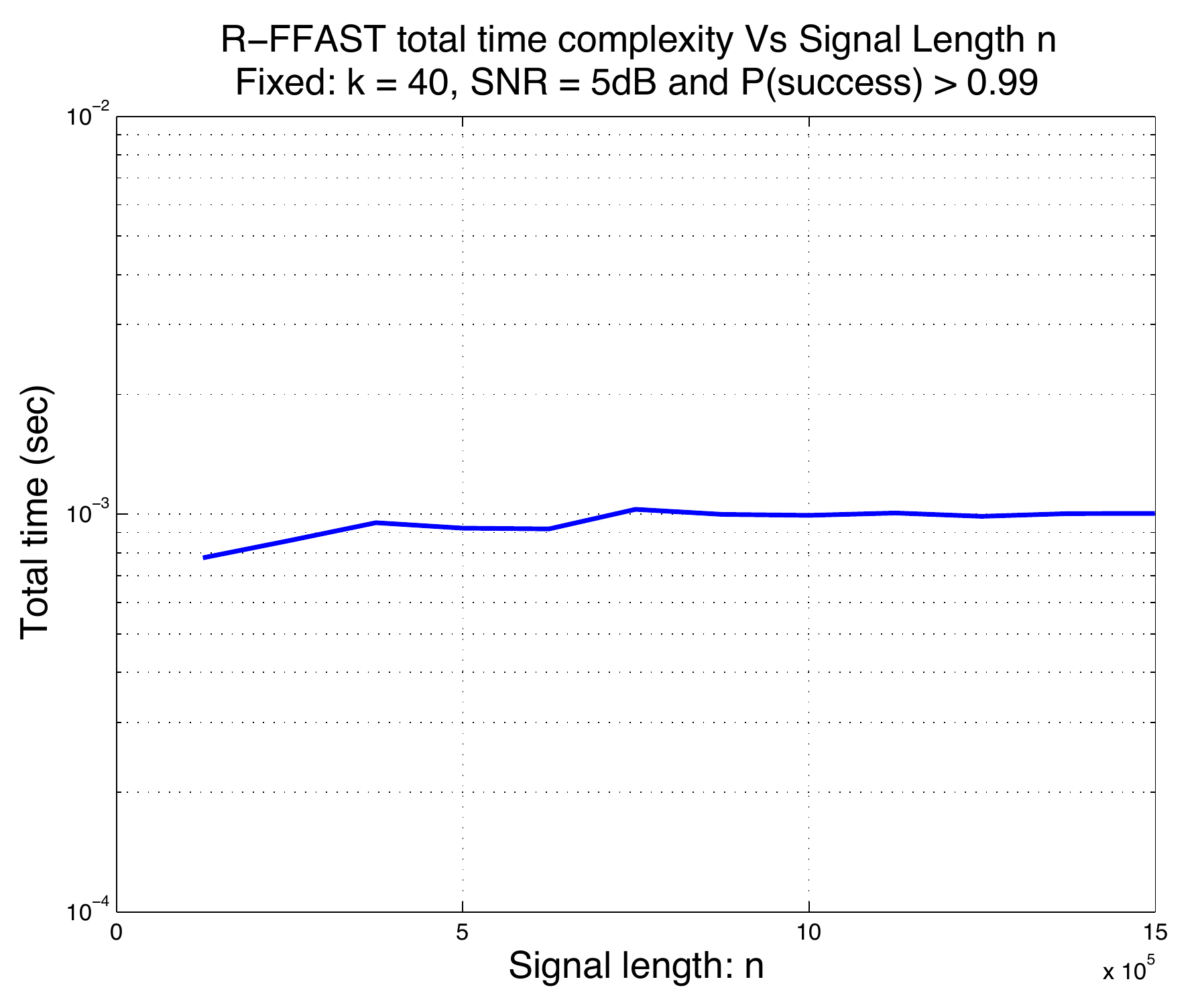}}
\caption{The plots show the scaling of the number of samples $\samples$ and the total time required by the R-FFAST algorithm to successfully reconstruct a $\sparsity=40$ sparse DFT $\transform$, for increasing signal length $\length$. For a fixed support recovery probability of $0.99$, sparsity $\sparsity$ and SNR of $5dB$, we note that both the {\em sample complexity} as well as {\em time complexity} of the R-FFAST scale {\em sub-linearly} in $\length$.}\label{fig:stvnplot}
\end{figure}

In \cite{pawar2013computing}, we have theoretically analyzed as well as empirically validated the scaling of the number of samples $\samples$ required by the FFAST algorithm, as a function of the sparsity $\sparsity$. In this section, we empirically evaluate the scaling of the number of samples $\samples$ and the processing time, required by the R-FFAST algorithm to successfully compute the DFT $\transform$, as a function of the ambient signal dimension $\length$, see Fig.~\ref{fig:stvnplot}.
\paragraph*{{\bf Simulation Setup}}
\begin{itemize}
\item An $\length$-length DFT vector $\transform$ with $\sparsity = 40$ non-zero coefficients is generated. The non-zero DFT coefficients are chosen to have a uniformly random support, fixed amplitude but an arbitrary random phase. The input signal $\signal$ is obtained by taking the IDFT of $\transform$. The length of the signal $\length$ is varied from $\length = 49*50*51 \approx 0.1$ million to $\length = 49*50*51*12 \approx 1.49$ million, i.e., $12\times$.
\item The noise-corrupted signal $\obs = \signal + \noise$, where $\noise$ is such that the effective signal-to-noise ratio is $5$dB, is processed through a $\stages =3$ stage R-FFAST architecture with $\delays = \samplesc\clusters$ delay-chains per stage. The sampling periods of the $3$ stages in the R-FFAST architecture are $50*51$, $51*49$ and $49*50$ respectively. This results in the total number of bins in the R-FFAST front-end architecture to be $\nbins = 150$. 
\item As the signal length $\length$ varies, the number of clusters $\clusters$ and the number of samples per cluster $\samplesc$ are varied, to achieve at least $0.99$ probability of successful support recovery.
\item Each sample point in the plot is generated by averaging over $1000$ runs. 
\end{itemize}

\subsubsection{R-FFAST sample complexity} In proposition~\ref{prop:DAccuracy}, we have shown that $\clusters = O(\log\length)$ clusters and $\samplesc = O(\log^{1/3}\length)$ samples per cluster are sufficient for decoding a singleton bin correctly using the bin-processing Algorithm~\ref{alg:binprocess}. In the empirical evaluation of the R-FFAST, we observed that $\samplesc = 3 \approx \log^{1/3}(\length)$ and $\clusters = [9:12] \approx \log(\length)$, were sufficient to successfully recover the support of the DFT $\transform$ with probability more than $0.99$. Thus, the overall empirical sample complexity $\samples \approx O(\sparsity\log^{4/3}\length)$, as shown in Fig.~\ref{fig:svnplot}. Note that, the empirical sample complexity performance coincides with the sample requirement for reliable decoding of a singleton-bin as pointed out by proposition~\ref{prop:DAccuracy} and is better than the theoretical claims of Theorem~\ref{thm:main}. We believe that this discrepancy is mainly due to the weaker analysis of a multi-ton bin error event in Appendix~\ref{app:main}. 

\subsubsection{R-FFAST time complexity} In Fig.~\ref{fig:tvnplot}, we plot the total time, sum of the time required for the front-end processing as well as the back-end peeling algorithm, required by the R-FFAST to successfully recover the DFT $\transform$, as a function of increasing signal length $\length$. Note that, a $12 \times$ fold increase in the ambient signal dimension $\length$ resulted in mere $30\%$ increase in the time complexity of the R-FFAST algorithm. Thus, empirically substantiating the claim of {\em sub-linear in signal length $\length$} time complexity of the R-FFAST algorithm.

\subsection{Application of the R-FFAST for MR imaging}\label{sec:sim_brain}
\begin{figure}[h!]
\centering     
\subfigure[Log intensity plot of the \mdft of the original `Brain' image. The red enclosed region is fully sampled and used for the stable inversion.]{\label{simfig:dftfull}\includegraphics[width=.3\linewidth]{brain_full_signal.png}}
\subfigure[Original `Brain' image in spatial domain.]{\label{simfig:spatialfull}\includegraphics[width=.3\linewidth]{brain_full_orig.jpg}}
\subfigure[Reconstructed `Brain' image using the $2D$ R-FFAST architecture along with the fully sampled center frequencies. The total number of Fourier samples used is $60.18\%$.]{\label{simfig:rfull}\includegraphics[width=.3\linewidth]{brain_full_peel.jpg}}
\subfigure[Log intensity plot of \mdft \ of the original `Brain' image, after application of the vertical difference operation.]{\label{simfig:dftdiff}\includegraphics[width=.3\linewidth]{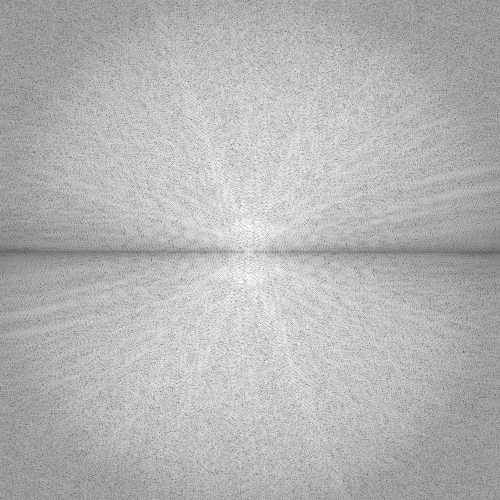}}
\subfigure[Differential `Brain' image obtained using the vertical difference operation on the original `Brain' image.]{\label{simfig:spatialdiff}\includegraphics[width=.3\linewidth]{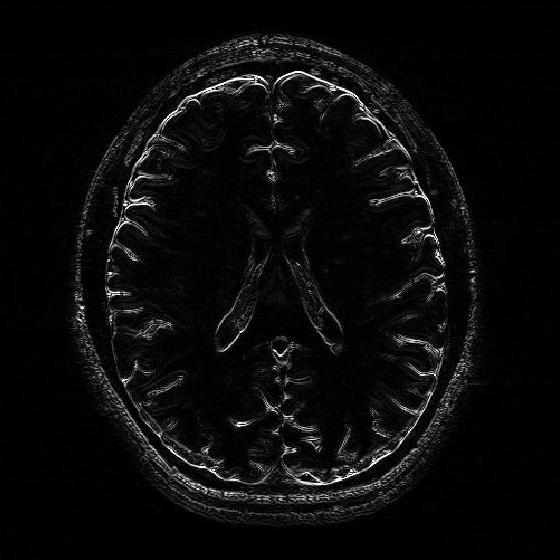}}
\subfigure[Differential `Brain' image reconstructed using the $2D$ R-FFAST architecture with $56.71\%$ of Fourier samples.]{\label{simfig:rdiff}\includegraphics[width=.3\linewidth]{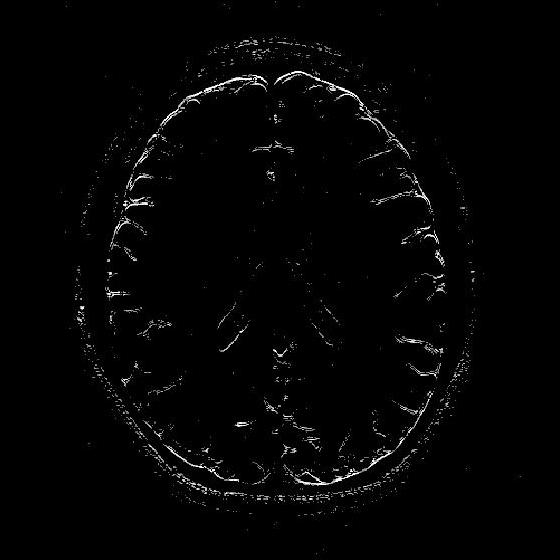}}
\caption{Application of the $2D$ R-FFAST algorithm to reconstruct the `Brain' image acquired on an MR scanner with dimension $504 \times 504$. We first reconstruct the differential `Brain' image shown in Fig.~\ref{simfig:spatialdiff}, using $\stages =3$ stage $2D$ R-FFAST architecture with $15$ random circular shifts in each of the $3$ stages of the R-FFAST architecture. Additionally we acquire all the Fourier samples from the center frequency as shown, by the red enclosure, in Fig.~\ref{simfig:dftfull}. Then, we do a stable inversion using the reconstructed differential `Brain' image of Fig.~\ref{simfig:rdiff} and the fully sampled center frequencies of Fig.~\ref{simfig:dftfull}, to get a reconstructed full `Brain' image as shown in Fig.~\ref{simfig:rfull}. Our proposed two-step acquisition and reconstruction procedure takes overall $60.18\%$ of Fourier samples.}
\end{figure}
The $1D$ R-FFAST architecture proposed in this paper can be generalized, in a straightforward manner, to $2D$ signals, with similar performance guarantees. In this section, we apply the $2D$ R-FFAST algorithm to reconstruct a brain image acquired on an MR scanner\footnote{The original MR scanner image is not pre-processed, other than retaining an image of size $504 \times 504$.} with dimension $504 \times 504$. In MR imaging, the samples are acquired in the Fourier domain and the task is to reconstruct the spatial image domain signal from significantly less number of Fourier samples. To reconstruct the full brain image using $2D$ R-FFAST, we perform the following two-step procedure: 
\begin{itemize}
\item {\em Differential space signal acquisition}: We perform a vertical finite difference operation on the image by multiplying the \mdft \ signal with $1 - e^{ 2 \pi \imath \omega_0}$. This operation effectively creates an approximately sparse differential image in spatial domain, as shown in Fig.~\ref{simfig:spatialdiff}, and can be reconstructed using R-FFAST. Note, that the finite difference operation can be performed on the sub-sampled data, and at no point do we need to access all the input Fourier samples. The differential brain image is then sub-sampled and reconstructed using a $3$-stage $2D$ R-FFAST architecture. Also, since the brain image is approximately sparse, we take $15$ delay-chains in each of the $3$ stages of the $2D$ R-FFAST architecture. The R-FFAST algorithm reconstructs the differential brain image using $56.71\%$ of Fourier samples.
\item {\em Inversion using fully sampled center frequencies}: After reconstructing the differential brain image, as shown in Fig.~\ref{simfig:rdiff}, we invert the finite difference operation by dividing the \mdft \ samples with $1 - e^{ 2 \pi \imath \omega_0}$. Since the inversion is not stable near the center of the Fourier domain, only the non-center frequencies are inverted and the center region is replaced by the fully sampled data.
\item Overall, the $2D$ R-FFAST algorithm uses a total of $60.18\%$ of Fourier samples to reconstruct the brain image shown in Fig.~\ref{simfig:rfull}.
\end{itemize}

%*********************************************************************************%
% Conclusion
%*********************************************************************************%
%\section{Conclusion and Future work}\label{ch5sec:conclusion}

%*********************************************************************************%
% Appendices
%*********************************************************************************%
\begin{appendices}
\section{Mutual incoherence bound}\label{app:maxcoherence}
Let $\svector{p}$ be a $p^{th}$ column of the bin-measurement matrix. Then, using the structure of $\svector{p}$ from \eqref{eq:steeringVec}, we have,
\begin{eqnarray}\label{eq:mumax}
\maxcoh(\mmatrix_{i,j}) &\leq& \max_{p \neq q} \frac{1}{\delays}|\svector{p}^{\dagger}\svector{q}| \nn\\
&=& \max_{\ell \neq 0} \frac{1}{\delays}|\sum_{s=0}^{\delays-1} \exp(\imath 2\pi\ell r_s/\length)| \nn\\
&=& \max_{\ell\neq 0}\coh(\ell) \ , 
\end{eqnarray}
where $\coh(\ell) \triangleq |\sum_{s=0}^{\delays-1} \exp(\imath 2\pi\ell r_s/\length)|/\delays$. 

Now consider the summation $\sum_{s=0}^{\delays-1} \cos(\imath 2\pi\ell r_s/\length)/\delays$ for any fixed $\ell \neq 0$. According to the hypothesis of the Lemma~\ref{lem:maxcoherence}, circular shifts $\{r_s\}_{s=0}^{\delays-1}$ are uniformly at random between $0$ and $\length-1$. Hence, the terms in the summation, i.e., $\{\cos(\imath 2\pi\ell r_s/\length)/\delays\}_{s=0}^{\delays-1}$, are i.i.d random variables with zero-mean and bounded support, given by $[-1/\delays,1/\delays]$. Using Hoeffding's inequality for the sum of independent random variables with bounded support we have, for any $t>0$,
\[
\pr(|\sum_{s=0}^{\delays-1} \cos(\imath 2\pi\ell r_s/\length)/\delays| > t) \leq 2\exp(-t^2\delays/2).
\]
Similarly,
\[
\pr(|\sum_{s=0}^{\delays-1} \sin(\imath 2\pi\ell r_s/\length)/\delays| > t) \leq 2\exp(-t^2\delays/2).
\]
Applying a union bound over the real and the imaginary parts of the summation term in $\coh(\ell)$, we get,
\begin{equation}
\pr(\coh(\ell) > \sqrt{2} t) \leq 4\exp(-t^2\delays/2).
\end{equation}

Further applying a union bound over all $\ell \neq 0$, we have,
\begin{eqnarray*}
\pr(\maxcoh(\mmatrix_{i,j}) > \sqrt{2} t) &\leq& 4\length\exp(-t^2\delays/2)\\
&=& 0.8
\end{eqnarray*}
for $t = \sqrt{2\log(5\length)/\delays}$. Thus, over all the random choices of the circular shifts $r_0,\hdots,r_{\delays-1}$, with probability at least 0.2,
\[ \maxcoh(\mmatrix_{i,j}) < 2\sqrt{\log(5\length)/\delays} \ , \ \ \forall \ i,j.\]
\eop

\section{Restricted-isometry-property}\label{app:RIP} 
Consider a $s$-sparse vector $\transform$ and a matrix $\mmatrix$. Using basic linear algebra we get the following inequality,
\begin{eqnarray*}
\lambda_{\min}(\mmatrix^{\dagger}\mmatrix) ||\transform||^2 \leq ||\mmatrix\transform||^2\leq \lambda_{\max}(\mmatrix^{\dagger}\mmatrix) ||\transform||^2
\end{eqnarray*}

The Gershgorin circle theorem \cite{gershgorin1931uber}, provides a bound on the eigen-values of a square matrix. It states that, every eigen-value of a square matrix lies within at least one of the Gershgoring discs. A Gershgorin disc is defined for each row of the square matrix, with diagonal entry as a center and the sum of the absolute values of the off-diagonal entries as radius. Note that by hypothesis of the Lemma~\ref{lem:RIP}, the diagonal entries of the matrix $\mmatrix^{\dagger}\mmatrix$ are equal to $1$. Hence, $\maxcoh(\mmatrix)$ provides an upper bound on the absolute values of the off-diagonal entries of $\mmatrix^{\dagger}\mmatrix$. 

Then, using the Gershgorin circle theorem we have,
\[
\left(1 - \maxcoh(\mmatrix)(s-1)\right)_{+} ||\transform||^2\leq ||\mmatrix\transform||^2 \leq \left(1 + \maxcoh(\mmatrix)(s-1)\right)||\transform||^2.
\]

\eop

\section{Proof of Theorem~\ref{thm:main}}\label{app:main} In this section we provide a proof of the main result of the paper. The proof consists of two parts. In the first part of the proof, we show that the R-FFAST algorithm reconstructs the DFT $\transform$, with probability at least $1 - O(1/\sparsity)$, using $\samples = O(\sparsity\log^3\length)$ samples. In the second part of the proof, we show that computational complexity of the R-FFAST decoder is $O(\sparsity\log^4\length)$.

\subsection{Reliability Analysis and sample complexity of the R-FFAST} 
\subsubsection{Reliability analysis}\label{sec:reliability}
Let $\eventf$ denote an event that the R-FFAST algorithm fails to reconstruct the DFT $\transform$. Let $\eventb$ be an event that the Algorithm~\ref{alg:binprocess}, after processing the observations of a bin, fails to correctly classify the bin as a zero-ton, a singleton or a multi-ton bin. In the case of a singleton bin successful decoding also involves identifying the support and the value of the contributing non-zero DFT coefficient correctly. Now, if we use $\eventsb$ to denote the event that during the entire decoding process there exists some bin for which Algorithm~\ref{alg:binprocess} make a wrong decision, then putting the pieces together, we get,
\begin{eqnarray}\label{eq:errorF}
\pr(\eventf) &<& \pr(\eventsb) + \pr(\eventf \mid \overline{\eventsb}) \nn\\
&\overset{(a)}{<}& \pr(\eventsb) + O(1/\sparsity)
\end{eqnarray}
where in $(a)$, we used the fact that when there are no bin-processing errors in the entire decoding procedure, i.e., event $\overline{\eventsb}$ occurs, the R-FFAST algorithm performs identical to the noiseless FFAST algorithm proposed in \cite{pawar2013computing}. In Section~\ref{sec:nbins}, we have shown that the R-FFAST front-end architecture, by construction, outputs $O(\sparsity)$ bins each with a vector observation. In \cite{pawar2013computing}, we show that the FFAST peeling-decoder processes each bin only constant number of times, i.e., peeling-iterations are constant. Thus, we bound the probability of the event $\eventsb$ using a union bound over the constant number of iterations required for the R-FFAST algorithm to reconstruct the DFT $\transform$ and over $O(\sparsity)$ bins used in the R-FFAST architecture to get,
\begin{eqnarray}\label{eq:errorE}
\pr(\eventsb) &<& \text{number of iterations} \times \text{number of bins} \times \pr(\eventb) \nn\\
&=& O(\sparsity)  \pr(\eventb) 
\end{eqnarray}
From \eqref{eq:errorF} and \eqref{eq:errorE}, we note that in order to complete the reliability analysis of the R-FFAST algorithm, we need to show that $\pr(\eventb) < O(1/\sparsity^2)$.

The following lemma plays a crucial role in the analysis of the event $\eventb$. It analyzes the performance of an energy-based threshold rule to detect the presence of an arbitrary valued complex vector $\vec{u}$, from its noise-corrupted samples.
\begin{lemma}\label{lem:energy} For an arbitrary valued complex vector $\vec{u} \in \complex^{\delays}$ and $\noise \sim \cnormal(0,\eye{\delays}{\delays})$, we have,
\begin{equation}
\pr(||\vec{u} + \noise||^2 < (1+\threshold)\delays) < \exp\left(-\frac{\delays(||\vec{u}||^2/\delays - \threshold)^2}{2 + 4||\vec{u}||^2/\delays}\right),
\end{equation}
for any constant $0 < \threshold < ||\vec{u}||^2/\delays$.
\begin{proof} Please see Appendix~\ref{app:energy}
\end{proof}
\end{lemma}

Without loss of generality, consider algorithm~\ref{alg:binprocess} processing the observations $\binobsvb$ of bin $b$. In Section~\ref{sec:ExFFAST}, using signal processing identities, we have shown that the bin observation noise $\nbinobsvb \sim  \cnormal(0,\eye{\delays}{\delays})$. The processed bin $b$ can either be a zero-ton bin, or a single-ton bin or a multi-ton bin. Next, we analyze the probability of success of the bin-processing algorithm~\ref{alg:binprocess} for all the three possibilities.
\paragraph{Analysis of a zero-ton bin} Consider a zero-ton bin $b$, with an observation $\binobsvb = \nbinobsvb$. Let $\eventz$ be an event that a zero-ton bin is not identified as a `zero-ton'. Then,
\begin{eqnarray}\label{eq:eventzbound}
\pr(\eventz) &=& \pr(||\nbinobsvb||^2 > (1+\threshold)\delays) \nn\\
&=& P(\chi^2_{2\delays} > 2(1+\threshold) \delays) \nn\\
&<& 2\exp(-\delays\threshold^2/9) \ \ \forall \ \ \threshold \in [0,1/3].
\end{eqnarray}
where the last inequality follows from a standard concentration bound for Lipschitz functions of Gaussian variables, along with the fact that the Euclidean norm is a $1$-Lipschitz function. Thus,
\begin{equation}\label{eq:eventz}
\pr(\eventz) < O(1/\sparsity^2), \ \ {\text{if}}  \ \delays = O(\log\sparsity) \ \text{and} \ \threshold \in [0,1/3].
\end{equation}

\paragraph{Analysis of a single-ton bin} Let $\binobsvb = \transformc{\ell}\svector{\ell} + \nbinobsvb$, be an observation vector of a  single-ton bin with contribution from non-zero DFT coefficient at support $\ell$ and value $\transformc{\ell}$. The steering vector $\svector{\ell}$ (see \eqref{eq:steeringVec} for detailed structure) is the $\ell^{th}$ column of the bin-measurement matrix. Let $\events$ be an event that a single-ton bin is not decoded correctly. The event $\events$ consists of the following three events.
\paragraph*{Single-ton bin is wrongly classified as a zero-ton bin}[$\eventsz$] Let $\eventsz$ denote an event that the single-ton bin fails the energy test of the Algorithm~\ref{alg:FFAST} and is classified as a zero-ton. 
\begin{eqnarray}\label{eq:eventsz}
\pr(\eventsz) &=& \pr(||\binobsvb||^2 < (1+\threshold)\delays) \nn\\
&=& \pr(||\transformc{\ell}\svector{\ell} + \nbinobsvb||^2 < (1+\threshold)\delays) \nn\\
&<& \exp\left(-\frac{\delays(\snr - \threshold)^2}{2 + 4\snr}\right) \ \ \ \forall \ \threshold < \snr
\end{eqnarray}
where the last inequality follows from using Lemma~\ref{lem:energy}, $\transformc{\ell} = \sqrt{\snr}e^{\imath\phi}$, and the norm of steering vector $||\svector{\ell}|| = \sqrt{\delays}$.

\paragraph*{Either the support or the value of the single-ton bin is detected wrongly}[$\eventss$] Let $\eventss$ denote an event that the Algorithm~\ref{alg:binprocess} wrongly detects either the support or the value of the singleton bin, i.e., 
\begin{eqnarray}\label{eq:support}
\pr(\eventss) &<& 2\pr(\hat{\ell} \neq \ell) + \pr(\hat{\transformc{\ell}} \neq \transformc{\ell} \mid \hat{\ell} = \ell)\nn\\
&<& O(1/\length^2) + \pr(\hat{\transformc{\ell}} \neq \transformc{\ell} \mid \hat{\ell} = \ell)
\end{eqnarray}
where last inequality follows from proposition~\ref{prop:DAccuracy} for $\delays = O(\log^{4/3}\length)$ carefully chosen samples per bin. 

From Algorithm~\ref{alg:binprocess}, we have $\hat{\transformc{\ell}} = \svector{\ell}^{\dagger}\binobsvb/\delays = \transformc{\ell} + \cnormal(0,1/\delays)$. Then, using the fact that non-zero DFT coefficients take value from a $\levels$-PSK constellation with magnitude $\sqrt{\snr}$ we have,
\begin{eqnarray}\label{eq:value}
\pr(\hat{\transformc{\ell}} \neq \transformc{\ell} \mid \hat{\ell} = \ell) &<& \pr(|\cnormal(0,1/\delays)| > \sqrt{\snr}\sin(\pi/\levels)) \nn\\
&=& \exp(- \delays\snr\sin^2(\pi/\levels)).
\end{eqnarray}

Hence, from \eqref{eq:support} and \eqref{eq:value} and using the fact that $\snr,\levels$ are constants, we have,
\begin{eqnarray}\label{eq:eventss}
\pr(\eventss) &<& O(1/\sparsity^2)
\end{eqnarray}
for $\delays = O(\log^{4/3}\length)$, carefully chosen samples as per discussion in Section~\ref{sec:delayCS}.

\paragraph*{Single-ton bin is wrongly classified as a multi-ton bin} [$\eventsm$] Let $\eventsm$ be an event that a singleton bin $b$ is processed by the Algorithm~\ref{alg:binprocess} and classified as a multi-ton bin. Thus,
\begin{eqnarray}\label{eq:eventsm}
\pr(\eventsm) &<& \pr(\eventsm \mid \overline{\eventss}) + \pr(\eventss) \nn\\
&=& \pr(\eventz) + \pr(\eventss)
\end{eqnarray}

Further using a union bound, we get an upper bound on the probability of event $\events$ as,
\begin{equation*}
\pr(\events) < \pr(\eventsz) + \pr(\eventss) + \pr(\eventsm   )
\end{equation*}

Thus, from \eqref{eq:eventz}, \eqref{eq:eventsz}, \eqref{eq:eventss} and \eqref{eq:eventsm}, we have,
\begin{equation}\label{eq:events}
\pr(\events) < O(1/\sparsity^2), \ \ {\text{if}} \ \delays = O(\log^{4/3}\length) \ \text{and} \ 0 < \threshold < \min\{1/3, \snr\}.
\end{equation}

\paragraph{Analysis of a multi-ton bin} 
Consider a multi-ton bin $b$, that has a contribution from $L \geq 2$ components. Then, the observation vector of this bin can be written as $\binobsvb= \mmatrix_b\vec{v} + \nbinobsvb$, where $\vec{v} \in \complex^{\length}$ is an $L$-sparse vector with $L$ non-zero DFT coefficients, $\mmatrix_b$ is the bin measurement matrix and $\nbinobsvb \sim \cnormal(0,\eye{\delays}{\delays})$. Let $\eventm$ be an event that a multi-ton bin is decoded as a single-ton bin with a support $\ell$ and the non-zero DFT coefficient $\transformc{\ell}$ for some $\ell$, i.e.,
\begin{equation}
\pr(\eventm) < \pr(||\mmatrix_{b}\vec{v} - \transformc{\ell}\svector{\ell} + \nbinobsvb||^2 < (1 + \threshold)\delays)
\end{equation}

First we compute a lower bound on the term $||\mmatrix_b\vec{v} - \transformc{\ell}\svector{\ell}||^2$ using the RIP Lemma~\ref{lem:RIP}. Let $\vec{u} = \vec{v} - \transformc{\ell}\vec{e}_{\ell}$, where $\vec{e}_{\ell}$ is a standard basis vector with $1$ at $\ell^{th}$ location and $0$ elsewhere. The vector $\vec{u}$ is either $L-1,L$ or $L+1$ sparse. Then, using the Lemmas  \ref{lem:maxcoherence}, \ref{lem:RIP} and the fact that all the non-zero components of $\vec{u}$ are constant, we have,
\begin{eqnarray}\label{eq:lblton}
||\mmatrix_{b}\vec{v} - \transformc{\ell}\svector{\ell}||^2 &=& ||\mmatrix_{b}\vec{u}||^2 \nn\\
&\geq& \constants{3}L\snr\delays(1 - \maxcoh(\mmatrix_{b})L)_{+} \nn\\
&>& \constants{3}L\snr\delays(1 - 2L\sqrt{\log(5\length)/\delays})_{+}
\end{eqnarray}
for some constant $\constants{3} > 0$.
Next, we compute an upper bound on the probability of the failure event $\eventm$,
\begin{eqnarray}\label{eq:boundeventm}
\pr(\eventm) &<& \pr(\eventm \mid L < \log(\sparsity)) + \pr(L \geq \log(\sparsity) ) \nn\\
&<& \pr(\eventm \mid L < \log(\sparsity)) + O(1/\sparsity^2),
\end{eqnarray}
where in last inequality we have used the fact that the number of the non-zero DFT coefficients connected to any bin is a Binomial $B(1/(\binexp\sparsity),\sparsity)$ random variable (see \cite{pawar2013computing} for more details), for some constant $\binexp > 0$. Hence to show that $\pr(\eventm) < O(1/\sparsity^2)$, we need to show $ \pr(\eventm \mid L < \log(\sparsity)) < O(1/\sparsity^2)$. Towards that end,
\begin{eqnarray*}
 \pr(\eventm \mid L < \log(\sparsity)) &=& \pr(||\mmatrix_{b}\vec{v} - \transformc{\ell}\svector{\ell} + \nbinobsvb||^2 < (1 + \threshold)\delays \mid L < \log(\sparsity))\\
&\overset{(a)} {<}& \max_{2\leq L < \log{\sparsity}} \exp\left(\frac{-\delays\left(\constants{3}L\snr\left(1 - 2L\sqrt{\log(5\length)/\delays}\right)_{+} - \threshold\right)^2}{2 + 4\constants{3}L\snr\left(1 - 2L\sqrt{\log(5\length)/\delays}\right)_{+}}\right),
\end{eqnarray*}
where in $(a)$ we used the Lemma~\ref{lem:energy} and the lower bound from \eqref{eq:lblton}.

Hence, 
\begin{equation}\label{eq:eventm}
\pr(\eventm) < O(1/\sparsity^2), \ \ {\text{if}} \ \delays = O(\log^{2}\sparsity\log\length),
\end{equation}
and $0 < \threshold < \constants{3}L\snr\left(1 - 2L\sqrt{\log(5\length)/\delays}\right)_{+}$, for $2 \leq L < \log{\sparsity}$.
\paragraph{Upper bound on the probability of event $\eventb$}
From \eqref{eq:eventz}, \eqref{eq:events} and \eqref{eq:eventm} we have,
\begin{equation}\label{eq:eventb}
\pr(\eventb) < O(1/\sparsity^2),
\end{equation}
for some constant $\threshold >0$, fixed $\snr > 0$ and $\delays = O(\log^2(\sparsity)\log(\length))$.

\subsection{Sample and Computational complexity of R-FFAST} 
In Section~\ref{sec:nbins}, we have shown that the R-FFAST front-end architecture, by construction, has a constant number of stages $\stages$ and outputs $O(\sparsity)$ bins. Additionally, in Section~\ref{sec:reliability}, we have shown that $\delays = O(\log^3\length)$, samples per bin are sufficient for reliable recovery of the DFT $\transform$. Hence, the total sample complexity of the R-FFAST algorithm in the presence of the observation noise is
\[
\samples = O(\sparsity\log^3\length).
\]

Moreover, in \cite{pawar2013computing}, we show that the FFAST peeling-decoder processes each bin only constant number of times, i.e., peeling-iterations are constant. Thus, the total computational cost of the R-FFAST algorithm is,
\begin{eqnarray}
&&\text{Total computational cost} \nn\\
&=& \text{cost of computing the front-end output} + \text{cost of back-end peeling} \nn\\
&=& O(\sparsity\log\sparsity) \times \delays \times \stages + \text{cost of back-end peeling} \nn\\
&<& O(\sparsity\log^4\length)+ \text{cost of back-end peeling} \nn\\
&\overset{(a)}=& O(\sparsity\log^4\length)+ O(\sparsity\log^{4/3}\length) \nn\\
&=& O(\sparsity\log^4\length),
\end{eqnarray}
where $(a)$ follows from proposition~\ref{prop:DAccuracy} and the fact that each bin is processed only constant number of times.
\eop

\section{Threshold based energy-detector}\label{app:energy}
In this section we provide a proof of the Lemma~\ref{lem:energy}. Let $\vec{u}\in \complex^{\delays}$ be an arbitrary complex $\delays$ dimensional vector and $\noise \in \cnormal(0,\eye{\delays}{\delays})$. Let $0 < \threshold < ||\vec{u}||^2/\delays$, be some constant. Then, using the tail bound for a non-central $\chi^2$ random variable (see equation (58b) in \cite{wainwright2009information}) we get,
\begin{eqnarray*}
&&\pr(||\vec{u} + \noise||^2 < (1+\threshold)\delays) \nn\\
&=& \pr(||\sqrt{2}\vec{u} + \sqrt{2}\noise||^2 < 2\delays(1+\threshold))\nn\\
&<& \exp\left(-\frac{\delays(||\vec{u}||^2/\delays - \threshold)^2}{2 + 4||\vec{u}||^2/\delays}\right)
\end{eqnarray*}
\eop

\section{Proof of proposition~\ref{prop:clusterAccuracy}}\label{app:clusterAccuracy}
Consider $\samplesc$ noise-corrupted time-domain samples of a single complex sinusoid, obtained by sampling the signal at a periodic interval of $1$, as follows,
\begin{eqnarray}
y(t) = A e^{j(\omega t + \phi)} + w(t), \ \ t = 0, 1, 2, \hdots, (\samplesc-1),
\end{eqnarray}
where the amplitude $A$, frequency $\omega$, and the phase $\phi$ are deterministic but unknown constants. The noise samples $w(t)\sim \cnormal(0,1)$. Then, as shown in \eqref{eq:kayEst} the MMSE estimate of the unknown frequency $\omega$ is given by,
\begin{equation}
\hat{\omega} = \left(\omega + \normal\left(0,\frac{6/\rho}{\samplesc(\samplesc^2-1)}\right)\right)_{2\pi},
\end{equation}
where $\rho = A^2$.
Then for any constant $\range$, there exists sufficiently large value of $\samplesc$ such that,
\begin{eqnarray}
\pr\left(|\omega - \hat{\omega}| > \frac{\pi}{\range}\right) &=& \pr\left(\left|\normal\left(0,\frac{6/\rho}{\samplesc(\samplesc^2-1)}\right)\right| > \frac{\pi}{\range}\right)\nn\\
&=& \pr\left(|\normal(0,1)| > \sqrt{\frac{\pi^2}{\range^2}\frac{\samplesc(\samplesc^2-1)}{6/\rho}}\right)\nn\\
&=& 2Q\left(\sqrt{\frac{\pi^2}{\range^2}\frac{\samplesc(\samplesc^2-1)}{6/\rho}}\right) \nn\\
&<& \exp\left(-\frac{\pi^2}{\range^2}\frac{\samplesc(\samplesc^2-1)}{6/\rho}\right) \nn\\
&<& 1/n^3,
\end{eqnarray}
where in the last inequality, we used $\samplesc = O(\log^{1/3}\length)$.

\section{Proof of proposition~\ref{prop:DAccuracy}}\label{app:DAccuracy}
Let $\omega_i$ be an estimate of $2^i\omega$ obtained using the observations of cluster $i$ as per estimation rule \eqref{eq:kayRule}. Also let $\Omega_i := (\omega_i -\pi/\range, \omega_i + \pi/\range)$, i.e., $|\Omega_i| = 2\pi/\range$. Define $E_i$ an event that $(2^i\omega)_{2\pi} \notin \Omega_i$, where $\omega$ is the true unknown frequency, of the form $2\pi\ell/n$, corresponding to the location $\ell$ of a singleton bin. Then, from proposition~\ref{prop:clusterAccuracy} we have $\pr(E_i) = \pr(E_0) < 1/\length^3$ for $\samplesc = O(\log^{1/3}\length)$. Applying a union bound over $\clusters$ clusters, we get, 
\begin{eqnarray}\label{eq:pfinal}
\pr(\omega \notin |\cap_{i=0}^{\clusters-1}\Omega_i/2^{i}|) &\leq& \sum_{i=0}^{\clusters-1} \pr(E_i)\nn\\ 
&<& \clusters/n^3\nn\\
&<& 1/n^2,
\end{eqnarray}
where in the last inequality we used the value of $\clusters = O(\log\length)$. 

From \eqref{eq:cintersect} we know,
\begin{eqnarray}\label{eq:exhrange}
|\cap_{i=0}^{\clusters-1}\Omega_i/2^{i}| &\leq& \frac{2\pi}{2^{\clusters-1}\range} \nn\\
&<& \frac{2\pi}{\length},
\end{eqnarray}
for appropriate choice of the constants. Hence using \eqref{eq:pfinal}, \eqref{eq:exhrange} and the fact that there are total of $n$ possible frequencies $2\pi\ell/n, \ell = 0,\hdots,n-1$, we conclude that the singleton-estimator algorithm identifies the true location of the non-zero coefficient of a singleton bin with probability at least $1 - O(1/\length^2)$.

\subsection{Computational Complexity} The computational complexity of processing samples of each cluster is $\samplesc = O(\log^{1/3}\length)$. There are total of $\clusters = O(\log\length)$ clusters. Hence, the overall complexity of the singleton estimator algorithm is $O(\samplesc\clusters) = O(\log^{4/3}\length)$.
\end{appendices}

%*********************************************************************************%
% Bibliography
%*********************************************************************************%
\bibliographystyle{IEEEtran}
\bibliography{thesis}

\begin{thebibliography}{10}
\providecommand{\url}[1]{#1}
\csname url@samestyle\endcsname
\providecommand{\newblock}{\relax}
\providecommand{\bibinfo}[2]{#2}
\providecommand{\BIBentrySTDinterwordspacing}{\spaceskip=0pt\relax}
\providecommand{\BIBentryALTinterwordstretchfactor}{4}
\providecommand{\BIBentryALTinterwordspacing}{\spaceskip=\fontdimen2\font plus
\BIBentryALTinterwordstretchfactor\fontdimen3\font minus
  \fontdimen4\font\relax}
\providecommand{\BIBforeignlanguage}[2]{{%
\expandafter\ifx\csname l@#1\endcsname\relax
\typeout{** WARNING: IEEEtran.bst: No hyphenation pattern has been}%
\typeout{** loaded for the language `#1'. Using the pattern for}%
\typeout{** the default language instead.}%
\else
\language=\csname l@#1\endcsname
\fi
#2}}
\providecommand{\BIBdecl}{\relax}
\BIBdecl

\bibitem{pawar2013computing}
S.~Pawar and K.~Ramchandran, ``Computing a k-sparse n-length discrete fourier
  transform using at most 4k samples and o (k log k) complexity,'' \emph{arXiv
  preprint arXiv:1305.0870}, 2013.

\bibitem{li2014sub}
X.~Li, S.~Pawar, and K.~Ramchandran, ``Sub-linear time support recovery for
  compressed sensing using sparse-graph codes,'' \emph{arXiv preprint
  arXiv:1412.7646}, 2014.

\bibitem{candes2005decoding}
E.~J. Candes and T.~Tao, ``Decoding by linear programming,'' \emph{IEEE
  Transactions on IT}, 2005.

\bibitem{prony1795essai}
R.~Prony, ``Essai experimental--,-,'' \emph{J. de l'Ecole Polytechnique}, 1795.

\bibitem{pisarenko1973retrieval}
V.~F. Pisarenko, ``The retrieval of harmonics from a covariance function,''
  \emph{Geophysical Journal of the Royal Astronomical Society}, vol.~33, no.~3,
  pp. 347--366, 1973.

\bibitem{schmidt1986multiple}
R.~Schmidt, ``Multiple emitter location and signal parameter estimation,''
  \emph{Antennas and Propagation, IEEE Transactions on}, 1986.

\bibitem{roy1989esprit}
R.~Roy and T.~Kailath, ``Esprit-estimation of signal parameters via rotational
  invariance techniques,'' \emph{Acoustics, Speech and Signal Processing, IEEE
  Transactions on}, vol.~37, no.~7, pp. 984--995, 1989.

\bibitem{donoho2006compressed}
D.~Donoho, ``Compressed sensing,'' \emph{Information Theory, IEEE Transactions
  on}, 2006.

\bibitem{candes2006near}
E.~Candes and T.~Tao, ``Near-optimal signal recovery from random projections:
  Universal encoding strategies?'' \emph{Information Theory, IEEE Transactions
  on}, vol.~52, no.~12, pp. 5406--5425, 2006.

\bibitem{candes2006robust}
E.~Cand{\`e}s, J.~Romberg, and T.~Tao, ``Robust uncertainty principles: Exact
  signal reconstruction from highly incomplete frequency information,''
  \emph{Information Theory, IEEE Transactions on}, vol.~52, no.~2, pp.
  489--509, 2006.

\bibitem{tropp2007signal}
J.~A. Tropp and A.~C. Gilbert, ``Signal recovery from random measurements via
  orthogonal matching pursuit,'' \emph{IEEE Transactions on IT}, 2007.

\bibitem{rauhut2012restricted}
H.~Rauhut, J.~Romberg, and J.~A. Tropp, ``Restricted isometries for partial
  random circulant matrices,'' \emph{Applied and Computational Harmonic
  Analysis}, 2012.

\bibitem{vetterli2002sampling}
M.~Vetterli, P.~Marziliano, and T.~Blu, ``Sampling signals with finite rate of
  innovation,'' \emph{Trans. on Sig. Proc.}, 2002.

\bibitem{dragotti2007sampling}
P.~Dragotti, M.~Vetterli, and T.~Blu, ``Sampling moments and reconstructing
  signals of finite rate of innovation: Shannon meets strang--fix,''
  \emph{Signal Processing, IEEE Transactions on}, vol.~55, no.~5, pp.
  1741--1757, 2007.

\bibitem{blu2008sparse}
T.~Blu, P.~Dragotti, M.~Vetterli, P.~Marziliano, and L.~Coulot, ``Sparse
  sampling of signal innovations,'' \emph{Signal Processing Magazine, IEEE},
  vol.~25, no.~2, pp. 31--40, 2008.

\bibitem{mishali2010theory}
M.~Mishali and Y.~Eldar, ``From theory to practice: Sub-nyquist sampling of
  sparse wideband analog signals,'' \emph{Selected Topics in Signal Processing,
  IEEE Journal of}, vol.~4, no.~2, pp. 375--391, 2010.

\bibitem{GGI02}
A.~C. Gilbert, S.~Guha, P.~Indyk, S.~Muthukrishnan, and M.~Strauss,
  ``Near-optimal sparse fourier representations via sampling,'' ser. STOC
  '02.\hskip 1em plus 0.5em minus 0.4em\relax New York, NY, USA: ACM, 2002.

\bibitem{gilbert2008tutorial}
A.~C. Gilbert, M.~J. Strauss, and J.~A. Tropp, ``A tutorial on fast fourier
  sampling,'' \emph{Signal Processing Magazine, IEEE}, 2008.

\bibitem{hassanieh2012nearly}
H.~Hassanieh, P.~Indyk, D.~Katabi, and E.~Price, ``Nearly optimal sparse
  fourier transform,'' in \emph{Proc. of the 44th SOTC}.\hskip 1em plus 0.5em
  minus 0.4em\relax ACM, 2012.

\bibitem{iwen2010combinatorial}
M.~Iwen, ``Combinatorial sublinear-time fourier algorithms,'' \emph{Foundations
  of Computational Mathematics}, vol.~10, no.~3, pp. 303--338, 2010.

\bibitem{ghazi2013sample}
B.~Ghazi, H.~Hassanieh, P.~Indyk, D.~Katabi, E.~Price, and L.~Shi,
  ``Sample-optimal average-case sparse fourier transform in two dimensions,''
  \emph{arXiv preprint arXiv:1303.1209}, 2013.

\bibitem{kay1989fast}
S.~Kay, ``A fast and accurate single frequency estimator,'' \emph{Acoustics,
  Speech and Signal Processing, IEEE Transactions on}, vol.~37, no.~12, pp.
  1987--1990, 1989.

\bibitem{tretter1985estimating}
S.~Tretter, ``Estimating the frequency of a noisy sinusoid by linear regression
  (corresp.),'' \emph{Information Theory, IEEE Transactions on}, vol.~31,
  no.~6, pp. 832--835, 1985.

\bibitem{candes2006stable}
E.~Candes, J.~Romberg, and T.~Tao, ``Stable signal recovery from incomplete and
  inaccurate measurements,'' \emph{Communications on pure and applied
  mathematics}, vol.~59, no.~8, pp. 1207--1223, 2006.

\bibitem{gershgorin1931uber}
S.~A. Gershgorin, ``Uber die abgrenzung der eigenwerte einer matrix,''
  \emph{Proceedings of the Russian Academy of Sciences. Mathematical series},
  pp. 749 -- 754, 1931.

\bibitem{wainwright2009information}
M.~J. Wainwright, ``Information-theoretic limits on sparsity recovery in the
  high-dimensional and noisy setting,'' \emph{Information Theory, IEEE
  Transactions on}, vol.~55, no.~12, pp. 5728--5741, 2009.

\end{thebibliography}
\end{document}